\documentclass[twocolumn,secnumarabic,amssymb, nobibnotes, aps, prd]{revtex4-2}

\usepackage[colorlinks=true, citecolor=blue, filecolor=magenta, unicode]{hyperref}
\usepackage{amsmath}
\usepackage{amsfonts}
\usepackage{amssymb}
\usepackage{graphicx}

\newcommand{\fref}[1]{Fig.~\ref{#1}}

\begin{document}

\title{Laplace's method for elastic scattering diagrams within multi-particle fields model}
\author{N.~Chudak}
\email{nata.podolyan@gmail.com}
\author{I.~Sharph}
\author{O.~Potiienko}
\author{A.~Mileva}
\author{K.~Merkotan}
\author{T.~Yushkevich}
\author{V.~Rusov}
\affiliation{Odessa Polytechnic State University, Shevchenko av., 1, Odessa, 65044, Ukraine}
\date{December 2021}%

\begin{abstract}
We apply the multi-particle fields model to calculate the differential cross-section $d\sigma/dt$ of elastic proton-proton scattering. This problem includes the calculation of multidimensional integrals arising from the loop Feynman diagrams. We demonstrated how these integrals can be reduced with Laplace's method to one- and two-dimensional integrals which can be calculated numerically. The obtained result qualitatively describe the minimum in differential cross-section dependency $d\sigma/dt(t)$.
\end{abstract}
\maketitle

\section{Introduction}
\label{intro}
The process of proton-proton elastic scattering is the subject of both theoretical and experimental research for the last few decades. The humanity have accumulated a lot of experimental data \cite{Cafagna:2020izf}, but we still not have a dynamic theory from the first principles \cite{Dremin:2017eng} yet. The descriptions of such processes are usually built with either the phenomenological approaches \cite{Csorgo:2018ruk} which are different variations of Regge theory \cite{Collins1977, Ryskin:2009qf}, or with the geometrical models \cite{Dremin:2019pza,Islam:2017muy,UgoAmaldi}. All these approaches are based on some assumptions that are not the corollaries of fundamental physical principles, but may even be in conflict with these principles. In particular, the Reggeized models \cite{FADIN197550,Kuraev:1977eng,Lipatov:2008eng} use the assumption that multi-Regge region \cite{KRFeng,MultiRggeBFKLeng,Duhr2019,Hentschinski:2020rfx} provides the dominant contribution to the integrals for observables. This region of the phase space contains the points corresponding to the significantly different values of energy-momentum of the secondary particles in the final state of the scattering \cite{Sharph:2011wm}, which violates the energy-momentum conservation law.

Among the various experimental data regarding the proton-proton(antiproton) scattering the most theoretical investigations have centred on the elastic scattering \cite{Dremin:2013eng}. The later is justified by expectation for the description of elastic processes to be significantly simpler compared to the description of inelastic ones. In particular, the scattering amplitude for elastic process is the function of just two Lorentz-invariants \cite{Collins1977}. The description of inelastic processes is then obtained from the description of elastic processes with the Abramovsky-Gribov-Kancheli (AGK) cutting rules \cite{AGKeng,Levin:1900tt,Bartels:2008ru,Pelicer:2018xzu}. However, this approach requires additional assumptions regarding the form of multi-Regge vertex \cite{Gribov_Reggeonnaja_technica_eng,BAKER19761} and introduction of unobservable quantity such as cross-section of scattering with an exchange of a certain number of Reggeons. All these assumptions are directed to establish the elastic scattering amplitude dependence on Mandelstam variable $s$. At the same time, the dependence on another variable $t$ remains undefined. In the simplest one-Reggeon model $t$-dependence is included in both Regge pole trejectory and residue. While the trajectory is defined by the masses of $t$-channel resonances, the residue remains completely undefined. In the context of Regge diagrams this residue is explained as the product of vertices corresponding to the interaction of Reggeon with the hadrons undergoing elastic scattering \cite{Nikitin1980,Shabelski:2014yba,Kaidalov:2003eng}. The certain $t$-dependence of such vertices is postulated in such a way that allows one to describe the experimentally measured differential cross-section ${d\sigma}/{dt}$ at low values of $t$. This dependence is usually postulated to be $ \exp \left( -{{R}^{2}}\left| t \right| \right) $, where $R$ is a fitted parameter also known as the Regge radius of hadron \cite{Nikitin1980}. As being said, such assumption provides the desirable $t$-dependence of the differential cross-section $d\sigma/dt$ of elastic scattering at low $t$ which is close to linear one in logarithmic scale applied to the cross-section axis \cite{Ter-Martirosyan}. However, this dependence is not linear on the whole range of measurements, but is non-monotonic and has maxima and minima \cite{NAGY1979221,PhysRevLett.54.2180,Antchev:2018edk,Abazov:2020rus}. The models with the multi-Regge exchange of simple poles and the model of quasi-eikonal multi-Regge vertices \cite{Ter-Martirosyan,Kaidalov:1983vn_eng, Kaidalov:1983ew} are failed to explain these features. Indeed, this non-monotonic behavior has been described within the additive quark model \cite{Shabelski:2014yba} due to the interference contributions from processes with an exchange of various number of Reggeons. As for the phenomenological approaches, this dependence is also reproduced within the models with a Regge multiple poles \cite{Dipol_eng,PhysRevD.87.114018,PhysRevD.76.074030}. However, these models have high uncertainty in $t$-dependence, since the increasing in the order of pole requires increasing in the number of terms in Laurent series whose forms are also need postulating.
 
The non-monotonic dependency of differential cross-section ${d\sigma}/{dt}(t)$ was also reproduced in the models with simple Regge poles of various signatures and with the two-Reggeon cuts \cite{refId0}.

The present research is devoted to the description of effect of non-monotonic $t$-dependence of the elastic scattering differential cross-section $d\sigma/dt$ built upon a purely dynamical model. In other words, we use the model that is based on the fundamental physical principles, beginning from the Lagrangian and its corresponding dynamical equations, quantization, and solution of the equations describing the dynamics of corresponding relativistic quantum system in Fock space. We do that in the framework of the multi-particle fields model.

\section{Problem statement and literature review}
We described the model of multi-particle fields in \cite{Korotca_statta_v_UJP, Ptashynskiy2019MultiparticleFO,Sharph:2020gki} and showed that hadrons can be considered as the quanta of field defined on the subsets of simultaneous events. These subsets are extracted from the tensor product of two (for mesons) and three (for barions) Minkowski spaces. The codomains of these fields are invariant subspaces of the tensor product of bispinors, where the scalar (for mesons) and bispinor (for protons) representations of Lorentz group act. The trivial representation of $SU_c\left(3\right)$ group acts also on these subsets, which corresponds to the fact that hadrons have no color. As usually, we build a Lagrangian for these fields and apply gauge principle. The local $SU_c\left(3\right)$-invariance of Lagrangian is provided by the gauge fields transforming by the tensor representation of the Lorentz group and internal symmetry groups. In this case, the common way of providing the gauge invariance through the introduction of covariant derivatives is the special case of the multi-particle approach for obtaining the local invariance. We have shown \cite{Korotca_statta_v_UJP, Ptashynskiy2019MultiparticleFO,Sharph:2020gki} that these tensor fields can be used for description of the creation and annihilation processes of glueballs, i.e. the bound states of confined gluons. The same fields provide both the confinement of quarks within the protons and mesons, and interaction between the quarks of different protons through the exchange of glueballs. As the result, we have the dynamical model for the three-particle bispinor fields that interact via the two-particle glueball fields. The calculation of observable quantities within this model can be performed with Feynman diagram technique, since such nonperturbative effects as the confinement of quarks and gluons are already accounted for in the internal dynamics of multi-particle field quanta. The non-zero masses of glueballs are also obtained from the dynamical equations in a natural way. This leads to the finite value of the elastic scattering amplitude at $t = 0$ due to the strong interaction. It is known that the total proton-(anti)proton scattering cross-section is finite after exclusion of electromagnetic interaction \cite{Cartiglia:2013jya,UgoAmaldi}. As the result of optical theorem \cite{Collins1977}, the scattering amplitude must have a finite value at $t = 0$. This finiteness is simply postulated in the mentioned phenomenological models. So it is necessary to develop a model that would describe the mentioned features of scattering amplitude without simply postulating them. In QCD perturbation theory one have an infrared singularity for the scattering with a massless gluon exchange. It is clearly problematic to use perturbative QCD at $t = 0$, where finiteness of the elastic scattering amplitude may be caused by nonperturbative effects. However, it is still unclear which nonperturbative effects and how they lead to the finiteness of the scattering amplitude. Meanwhile, in the model of multi-particle fields it is the non-zero mass of the glueball (which is actually the consequence of nonperturbative effects) that leads to the finite value of the scattering amplitude.

\begin{figure}[tbp!]
	\centering
	\includegraphics[width=1\linewidth]{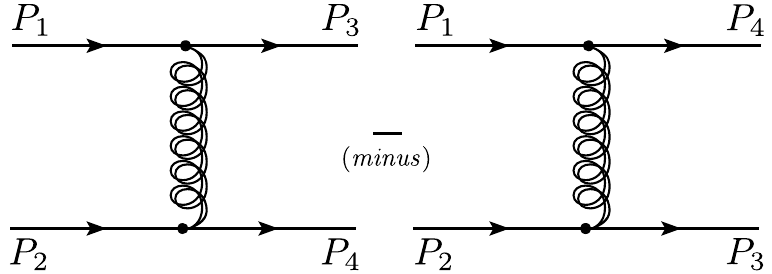}
	\caption{
	Pole tree-level diagrams of elastic proton-proton scattering. ${{P}_{1}},{{P}_{2}}$ are the four-momenta of initial protons, ${{P}_{3}},{{P}_{4}}$ are the four-momenta of outgoing protons. Double lines correspond to the glueballs (bound state of gluons). The \textit{minus} between the diagrams reflects the fact that protons are described by Fermi-Dirac statistics.
	}
	\label{fig:polusnaya}
\end{figure}

The qualitative description of the inelastic scattering may be done considering tree-level diagrams only \cite{PhysRevD.101.076021}. However, the tree-level diagrams (see \fref{fig:polusnaya}) are not enough to describe the elastic scattering. Nevertheless, the non-monotonicity of the cross-section dependence appears even in such a simple model, except that it reproduces only the minimum, but not maximum.

Therefore, the description of experimental data requires the calculation of more complex loop diagrams.

The calculation of such diagrams is reduced to the calculation of multidimensional integral over virtual four-momenta. 
Similar integrals have been considered in \cite{Eden:98637} with an application of Feynman parametrization, where, however, only the location of the singularities was studied but not the calculation method. The Feynman identity simplifies the integrand due to the fact that instead of the product of multiple Feynman denominators we obtain the power of single denominator, while, unfortunately, it also leads to the essentially more complex integration domain. For this reason, here we consider an approximate method of calculation starting from the expression for the loop diagram contribution to the elastic scattering amplitude. There is also another approach to the similar integrals \cite{PhysRev.186.1656} based on the eikonal approximation. However, such approximation significantly changes the pole structure of the integrand. While the eikonal approximation allows one to reduce the dimension of the integral for the eikonal, such approach does not solve the problem of calculating the integrals themselves. The main problem here is to calculate the limit of multidimensional integral as the parameters that shift the integrand poles tend to zero. Note that taking the limit inside the integral is not allowed in this case, because the poles will move inside the integration domain so the integral will diverge. This puts restrictions on the application of numerical integration methods, because taking the limit in the end requires the calculation at small parameters, which makes the poles close to the integration domain and thus complicates the numerical calculations.

To solve the outlined problem, we use the Laplace's method \cite{DeBruijn:225131} which worked well for the description of inelastic scattering processes \cite{Sharf:2011ujp,Sharf:2006it, Sharf:2007cf, Sharf:2009yy}. This method allowed us to effectively calculate the integrals with dimension up to 100. However, in those works the limit could be taken before the integration, because the poles of integrands were outside of the integration domain, which is not the case here. 

Now let us figure out how to apply Laplace's method in the present situation. We know that the problem arises from the Feynman denominators that correspond to the lines of a diagram. Let us number these lines in an arbitrary order and put the expression of each denominator in the form ${{\left( {{z}_{a}}-i\varepsilon  \right)}^{-1}}$, where $a$ is the line number, $z_a$ is the expression corresponding to the $a$-th line, $ \varepsilon $ is the parameter which should be made zero after the integral is calculated. Selecting a subset of $k$ lines we can equate the corresponding expressions $z_a$ to zero and consider the obtained system of equations. If the obtained system of equations is consistent, it defines the subset of integration domain where the integrand is equal to ${{\varepsilon }^{-k}}$. As a result, the major contribution to the integral comes from the region in which the greatest number of denominators is equal to zero. Let us denote this number by $l$ and the total number of integration variables by $n$. From the corresponding system of equations we can express the $l$ variables through the rest $n - l$ ones. Then it is convenient to change the first $l$ integration variables. These new variables are the deviations from the values that satisfy the equation system for $l$ denominators. It means that the absolute value of the integrand now has a distinct maximum at zero values of the first $l$ variables regardless of the next $n - l$ variables. Then we can apply the Laplace's method to integrate over the first $l$ variables. As we have already mentioned, the absolute value of the integrand at the maximum point is equal to ${{\varepsilon }^{-l}}$. At the same time, when applying Laplace's method, there comes the gaussian integral which leads to the factor ${{\varepsilon }^{-l}}$. It removes the $\varepsilon$ from expression and allows one to turn $\varepsilon$ to zero before the integration. Then the obtained integral can be calculated using numerical methods.

\begin{figure}[tbp!] 
	\includegraphics[width=1\linewidth]{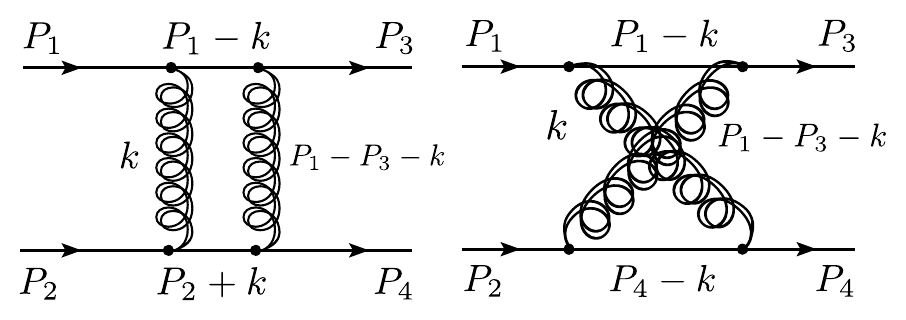}
	\caption{\label{fig:simplest_loops} The simplest loop diagrams for the elastic scattering of protons. ${{P}_{1}},{{P}_{2}}$ and ${{P}_{3}},{{P}_{4}}$ are the four-momenta of the incoming and outgoing protons correspondingly, $k$ is the four-momentum of the virtual particle. Double gluon lines correspond to the glueballs (bound state of gluons).}
\end{figure}

In the present paper we apply the described idea to calculate the sum of simplest single-loop diagrams for the elastic scattering of protons (\fref{fig:simplest_loops}).

We calculate the elastic proton-proton scattering differential cross-section within the multi-particle fields approach. We consider the contributions of the diagrams in Figs.~\ref{fig:polusnaya},~\ref{fig:simplest_loops}, including the diagrams obtained from those depicted on \fref{fig:simplest_loops} by interchanging the final-state particle lines. Finally, we compare the calculation results with the experimental data.

\section{Laplace's method and passage to the limit $\varepsilon \to 0$}

An analytical expression for the diagram in \fref{fig:simplest_loops}a has the following form 
\begin{equation}\label{vyraz_diagramma33}
\begin{split}	
 A=\frac{{{\left( ig \right)}^{4}}}{{{\left( 2\pi  \right)}^{6}}}&{{\left( \bar{v}_{{{v}_{3}}}^{+}\left( {{P}_{4}} \right) \right)}_{{{s}_{4}}}}\gamma _{{{s}_{4}}{{s}_{2}}}^{a}{{\left( v_{{{v}_{2}}}^{-}\left( {{P}_{2}} \right) \right)}_{{{s}_{2}}}} \\
 \times &{{\left( \bar{v}_{{{v}_{4}}}^{+}\left( {{P}_{3}} \right) \right)}_{{{s}_{3}}}}\gamma _{{{s}_{3}}{{s}_{1}}}^{b}{{\left( v_{{{v}_{1}}}^{-}\left( {{P}_{1}} \right) \right)}_{{{s}_{1}}}}\\
\times & \delta \left( \left( {{P}_{3}}+{{P}_{4}} \right)-\left( {{P}_{1}}+{{P}_{2}} \right) \right) t_{ab},
\end{split}
\end{equation}
\noindent
where $g$ is the effective coupling, ${{M}_{p}}$ and ${{M}_{G}}$ are the masses of the proton and the glueball respectively,  
${{\left( \bar{v}_{{{v}_{3}}}^{+}\left( {{P}_{4}} \right) \right)}_{{{s}_{4}}}},{{\left( v_{{{v}_{2}}}^{-}\left( {{P}_{2}} \right) \right)}_{{{s}_{2}}}}, {{\left( \bar{v}_{{{v}_{4}}}^{+}\left( {{P}_{3}} \right) \right)}_{{{s}_{3}}}}, {{\left( v_{{{v}_{1}}}^{-}\left( {{P}_{1}} \right) \right)}_{{{s}_{1}}}}$ - are the solutions of the Dirac equations, $\gamma _{{{s}_{4}}{{s}_{2}}}^{a},\gamma _{{{s}_{3}}{{s}_{1}}}^{b}$ - are the elements of the Dirac matrices, and $t_{ab}$ - is the tensor whose components are defined by
\begin{equation}\label{tensor_ab}
\begin{split}	
& {{t}_{ab}}=\int{{{d}^{4}}}k\left( {{k}_{a}}+2{\left(P_2\right)_a} \right)\left( 2{\left(P_1\right)_b}-{{k}_{b}} \right)
  \\ 
& \times \frac{1}{M_{p}^{2}-{{\left( {{P}_{1}}-k \right)}^{2}}-i\varepsilon }\frac{1}{M_{p}^{2}-{{\left( {{P}_{2}}+k \right)}^{2}}-i\varepsilon } \\ 
& \times \frac{1}{M_{G}^{2}-{{k}^{2}}-i\varepsilon }\frac{1}{M_{G}^{2}-{{\left( {{P}_{1}}-{{P}_{3}}-k \right)}^{2}}-i\varepsilon },
\end{split}
\end{equation}
\noindent
where $a, b = 0 .. 3$.

We consider the problem in the center of mass reference frame with the right hand coordinate system whose $z$ axis is oriented along ${{\vec{P}}_{1}}$, and the $x$ axis is perpendicular to $z$ and lies the plane containing vectors ${{\vec{P}}_{1}}$ and ${{\vec{P}}_{3}}$. All quantities are expressed in the units of the proton mass, ${{M}_{G}}$ and $g$ are considered as the model parameters.

First, we need to calculate the components of the tensor ${{t}_{ab}}$. After the tensor components $t_{ab}$ have been calculated, we use the squared absolute value of the scattering amplitude to calculate the elastic scattering differential cross-section $d\sigma/dt$ taking into account the contributions from the pole diagram from \fref{fig:polusnaya}.

To calculate the integral in \eqref{tensor_ab} we use Laplace's method. According to the calculation method described in the previous section, for a small value of $\varepsilon$ we determine which region makes the major contribution to the integral. If the $\varepsilon$ is small and non-zero, the absolute value of the integrand reaches its maximum in a region where the maximal number of real parts of the denominators (i.e. the parts that do not contain $\varepsilon$) turn to zero. Let's denote the real parts of the denominators in  \eqref{tensor_ab} as
\begin{equation}\label{znamenyky}
\begin{aligned}[c]
& {{z}_{1}}=M_{p}^{2}-{{\left( {{P}_{1}}-k \right)}^{2}} \\ 
& {{z}_{2}}=M_{p}^{2}-{{\left( {{P}_{2}}+k \right)}^{2}} \\ 
& {{z}_{3}}=M_{G}^{2}-{{k}^{2}} \\
& {{z}_{4}}=M_{G}^{2}-{{\left( {{P}_{1}}-{{P}_{3}}-k \right)}^{2}}. 
\end{aligned}
\end{equation}
Taking into account \eqref{znamenyky}, expression \eqref{tensor_ab} can be rewritten as
\begin{equation}\label{tensor_ab_short}
\begin{split}	
& {{t}_{ab}}=\int{{{d}^{4}}}k 
{{f}_{ab}}\left( {{k}_{0}},\vec{k} \right)
\prod_{j=1}^4\frac{1}{\left(z_j - i\varepsilon\right)}.
\end{split}
\end{equation}


The squared absolute value of the denominator in \eqref{tensor_ab} is the sum of the real part squared and the $\varepsilon^2$. When the real part in \eqref{znamenyky} is equal to zero and $\varepsilon$ is non-zero, then the integral converges and has the maximum. Assume that one of the denominators \eqref{znamenyky} is zero, which yields some subset of the integration domain where the real part of the denominator is zero. Again, considering the integration over this subset, the main contribution will be provided by that region where the real parts of some of the other denominators turn to zero. So a natural question to ask is how many expressions in \eqref{znamenyky} can be turned to zero at the same time? It has been shown \cite{sharph2020laplace} that either the first pair of expressions ${{z}_{1}}$ and ${{z}_{2}}$ corresponding to the 
horizontal lines in Feynman diagram (\fref{fig:simplest_loops}a), or another pair ${{z}_{3}}$ and ${{z}_{4}}$ corresponding to the vertical lines, can be turned to zero simultaneously. Thus, both the horizontal and vertical lines cannot be turned to zero at the same time. Next we present the calculations of each denominator in more detail.

Let us consider the tensor \eqref{tensor_ab} and denote the numerator as  ${{f}_{ab}}\left( {{k}_{0}},\vec{k} \right)=\left( {{k}_{a}}+2{{P}_{2a}} \right)\left( 2{{P}_{1b}}-{{k}_{b}} \right)$. Taking into account that in the center of mass reference frame ${{\vec{P}}_{2}}=-{{\vec{P}}_{1}}$, the denominator $z_2$ in the tensor \eqref{tensor_ab} may be represented as follows

\begin{equation}\label{z2_cm}
{{z}_{2}}= M_{p}^{2}-{{\left( \frac{\sqrt{s}}{2}-{{k}^{0}} \right)}^{2}}+{{\left( \vec{k}-{{{\vec{P}}}_{1}} \right)}^{2}}.
\end{equation}

\begin{figure}[tbp!]
	\centering
	\includegraphics[width=0.95\linewidth]{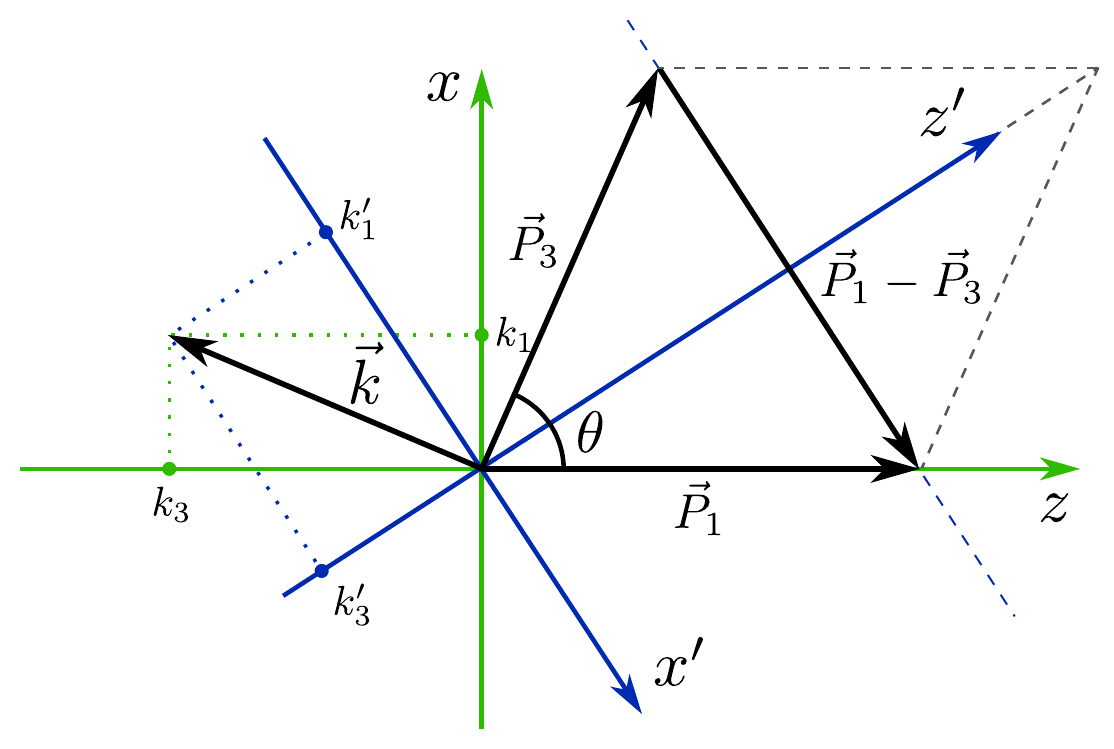}
	\caption{
	The coordinates ${{k}^{1}}$ and ${{k}^{3}}$ of the vector $\vec{k}$ in initial coodrinate system and its coordinates ${{{k}'}^{1}}$ and ${{{k}'}^{3}}$ after transformation of the coordinate system.}
	\label{fig:osi_1}
\end{figure}

Tensor \eqref{tensor_ab_short} contains now 7 non-zero terms, and each of them has to be calculated separately. Note that they differ only in numerators and have the same denominators, which means they all can be calculated in the same way. It is then convenient to change the coordinate system as shown in \fref{fig:osi_1}.

The vector $\vec{k}$ has the coordinates ${{k}^{1}}$ along $x$ axis and ${{k}^{3}}$ along $z$ axis. We transform the axes so that ${x}'$ becomes parallel to the vector ${{\vec{P}}_{1}}-{{\vec{P}}_{3}}$. The coordinates of vector $\vec{k}$ expressed through the new coordinates ${{{k}'}^{1}}$ and ${{{k}'}^{3}}$ in the transformed system are
\begin{equation}\label{k1k3}
\begin{split}
& {{k}^{1}}=-{{{{k}'}}^{1}}\cos \left( \frac{\theta }{2} \right)+{{{{k}'}}^{3}}\sin \left( \frac{\theta }{2} \right), \\ 
& {{k}^{3}}={{{{k}'}}^{1}}\sin \left( \frac{\theta }{2} \right)+{{{{k}'}}^{3}}\cos \left( \frac{\theta }{2} \right). \\ 
\end{split}
\end{equation}  

The numerator ${{f}_{ab}}\left( {{k}_{0}},\vec{k} \right)=\left( {{k}_{a}}+2{{P}_{2a}} \right)\left( 2{{P}_{1b}}-{{k}_{b}} \right)$  in \eqref{tensor_ab_short} is also expressed through the coordinates \eqref{k1k3}. For a further calculation of \eqref{tensor_ab_short} we also change the variables as follows
\begin{equation}\label{zamina}
\begin{aligned}[c]
{{q}^{0}} = {{k}^{0}}
\end{aligned}
\qquad \qquad
\begin{aligned}[c]
\vec{q} = \frac{{{{\vec{P}}}_{1}}-{{{\vec{P}}}_{3}}}{2}-\vec{k},
\end{aligned}
\end{equation}
\noindent
so the tensor \eqref{tensor_ab_short} can be rewritten as
\begin{equation}\label{tensor_t_through_q}
\begin{split}
& {{t}_{ab}}=\int\limits_{-\infty }^{\infty }{d{{q}^{0}}}\int{d\vec{q}}{{f}_{ab}}\left( {{q}^{0}},\vec{q} \right)
\prod_{j=1}^4\frac{1}{\left(z_j - i\varepsilon\right)},
\end{split}
\end{equation}
where the denominators \eqref{znamenyky} take the following form
\begin{widetext}
\begin{equation}\label{z_through_q}
\begin{split}
& z_1 = M_{p}^{2}-{{\left( \frac{\sqrt{s}}{2}-{{q}^{0}} \right)}^{2}}+{{\left( {{q}^{1}} \right)}^{2}}+{{\left( {{q}^{2}} \right)}^{2}}+{{\left( \frac{\left| {{{\vec{P}}}_{1}}+{{{\vec{P}}}_{3}} \right|}{2}+{{q}^{3}} \right)}^{2}} \\
& z_2 = M_{p}^{2}-{{\left( \frac{\sqrt{s}}{2}+{{q}^{0}} \right)}^{2}}+{{\left( {{q}^{1}} \right)}^{2}}+{{\left( {{q}^{2}} \right)}^{2}}+{{\left( \frac{\left| {{{\vec{P}}}_{1}}+{{{\vec{P}}}_{3}} \right|}{2}+{{q}^{3}} \right)}^{2}} \\
& z_3 = M_{G}^{2}-{{\left( {{q}^{0}} \right)}^{2}}+{{\left( \frac{\left| {{{\vec{P}}}_{1}}-{{{\vec{P}}}_{3}} \right|}{2}-{{q}^{1}} \right)}^{2}}+{{\left( {{q}^{2}} \right)}^{2}}+{{\left( {{q}^{3}} \right)}^{2}} \\
& z_4 = M_{G}^{2}-{{\left( {{q}^{0}} \right)}^{2}}+{{\left( \frac{\left| {{{\vec{P}}}_{1}}-{{{\vec{P}}}_{3}} \right|}{2}+{{q}^{1}} \right)}^{2}}+{{\left( {{q}^{2}} \right)}^{2}}+{{\left( {{q}^{3}} \right)}^{2}}.
\end{split}
\end{equation} 
\end{widetext}

Note that $z_1$ and $z_2$ differ only in sign before $q^0$. At the same time the only difference between $z_3$ and $z_4$ is the sign before $q^1$. It allows us to shorten the calculations by introducing the following notations
\begin{equation}\label{zpm}
\begin{split}
& z_1^{\pm}=
M_{p}^{2}-{{\left( \frac{\sqrt{s}}{2}\pm{{q}^{0}} \right)}^{2}}
+{{\left( {{q}^{1}} \right)}^{2}}
+{{\left( {{q}^{2}} \right)}^{2}}
\\
& \hphantom{\hspace{3cm}}
+{{\left( \frac{\left| {{{\vec{P}}}_{1}}+{{{\vec{P}}}_{3}} \right|}{2}+{{q}^{3}} \right)}^{2}} 
\\
& z_3^{\pm} = 
M_{G}^{2}-{{\left( {{q}^{0}} \right)}^{2}}
+{{\left( \frac{\left| {{{\vec{P}}}_{1}}-{{{\vec{P}}}_{3}} \right|}{2} \pm {{q}^{1}} \right)}^{2}}
\\
& \hphantom{\hspace{4cm}}
+{{\left( {{q}^{2}} \right)}^{2}}+{{\left( {{q}^{3}} \right)}^{2}}, 
\end{split}
\end{equation} 
\noindent
so that $z_1 = z_1^-$, $z_2 = z_1^+$, and $z_3 = z_3^-$, $z_4 = z_3^+$.

Since the differential cross-section depends on the transmitted four-momentum $t$, we express the vectors ${{\vec{P}}_{1}}$ and ${{\vec{P}}_{3}}$ through Mandelstam variable $t$
\begin{equation}\label{cherez_t}
\begin{split}
& \left| {{{\vec{P}}}_{1}}-{{{\vec{P}}}_{3}} \right|=\sqrt{\left| t \right|}
\\
& \left| {{{\vec{P}}}_{1}}+{{{\vec{P}}}_{3}} \right|=\sqrt{4{{P}^{2}}-\left| t \right|},
\end{split}
\end{equation}
\noindent
where $P = | \vec{P}_1 | = | \vec{P}_3 |$.

Applying one more change of variable $${{q}^{3}} \rightarrow \frac{1}{2}\sqrt{4{{P}^{2}}-\left| t \right|}+{{q}^{3}}$$ and taking into account \eqref{cherez_t}, we obtain the new expressions for \eqref{zpm}:
\begin{equation}\label{zpm_after_q3_change}
\begin{split}
& z_1^{\pm} = M_{p}^{2}-{{\left( \frac{\sqrt{s}}{2} \pm {{q}^{0}} \right)}^{2}}+{{\left( {{q}^{1}} \right)}^{2}}+{{\left( {{q}^{2}} \right)}^{2}}+{{\left( {{q}^{3}} \right)}^{2}}
\\
& z_3^{\pm} = M_{G}^{2}-{{\left( {{q}^{0}} \right)}^{2}}+{{\left( \frac{\sqrt{\left| t \right|}}{2} \pm {{q}^{1}} \right)}^{2}}+{{\left( {{q}^{2}} \right)}^{2}}
\\ 
&\hphantom{\hspace{3cm}}+ {{\left( {{q}^{3}}-\frac{\sqrt{4{{P}^{2}}-\left| t \right|}}{2} \right)}^{2}}
. 
\end{split}
\end{equation} 

Finally, introducing the spherical coordinates $\left( q, \theta, \phi \right)$:
\begin{equation}\label{sferychni_koordynaty}
\begin{split}
& {{q}^{1}}=q\sin \left( \theta  \right)\cos \left( \varphi  \right), \\ 
& {{q}^{2}}=q\sin \left( \theta  \right)\sin \left( \varphi  \right), \\ 
& {{q}^{3}}=q\cos \left( \theta  \right),  
\end{split}
\end{equation}
\noindent
where $0 \leq \theta \leq \pi$ and $0 \leq \phi < 2\pi$, we can rewrite \eqref{tensor_t_through_q} as follows
\begin{equation}\label{tensor_in_sphere}
\begin{split}
 {{t}_{ab}}=&\int\limits_{-\infty }^{\infty }{d{{q}^{0}}}\int\limits_{0}^{\infty }{{{q}^{2}}dq}\int\limits_{0}^{\pi }{\sin \left( \theta  \right)}d\theta \int\limits_{0}^{2\pi }{d\varphi }{{f}_{ab}}\left( {{q}^{0}},q,\theta ,\varphi  \right) \\
& \times 
\frac{1}{\left(z_1^{+} - i\varepsilon\right)}
\frac{1}{\left(z_1^{-} - i\varepsilon\right)}
\frac{1}{\left(z_3^{-} - i\varepsilon\right)}
\frac{1}{\left(z_3^{+} - i\varepsilon\right)},
\end{split}
\end{equation}
\noindent
where 
\begin{equation}\label{zpm_in_sphere}
\begin{split}
& z_1^{\pm} = M_{p}^{2}-{{\left( \frac{\sqrt{s}}{2} \pm {{q}^{0}} \right)}^{2}}+{{q}^{2}}
\\
&  z_3^{\pm} = M_{G}^{2}-{{\left( {{q}^{0}} \right)}^{2}}
+ {\left(\frac{\sqrt{\left| t \right|}}{2} \mp q\sin \left( \theta  \right)\cos \left( \varphi  \right)\right)}^{2} \\
& + {\left( q\sin \left( \theta  \right)\sin \left( \varphi  \right) \right)}^{2}
+ {\left( q\cos \left( \theta  \right)-\frac{\sqrt{4{{P}^{2}}-\left| t \right|}}{2} \right)}^{2}
\end{split}
\end{equation}

As mentioned in the problem statement, we are going to use the Laplace's method to calculate the multidimensional integral \eqref{tensor_in_sphere}.

 The essential idea behind this method is that the integral whose integrand has a single maximum point in the integration domain, can be approximated nicely by the corresponding Gaussian integral. We have already seen above that either the first pair of denominators (associated with the horizontal lines of the diagram in \fref{fig:simplest_loops}a) or the second pair (two vertical lines) can be turned to zero simultaneously. Thus, we examine the first and the second pairs of denominators separately. We find the corresponding regions of the integration domain where either the first or the second pair of denominators turn into zero, and calculate the contribution of these regions to the integral.


\subsection{The first pair of denominators} \label{first_pair_section}
Let us consider the system of equations for the first two denominators $\left(z_1^{+} - i\varepsilon\right)$ and $\left(z_1^{-} - i\varepsilon\right)$ in \eqref{tensor_in_sphere}. If we set the real parts of these denominators to zero, their product reduces to ${\varepsilon }^{2}$, which provides the maximal contribution to the integrand in \eqref{tensor_in_sphere}.
\begin{equation}\label{systema_rivnan}
\left\{ \begin{split}
& z_1^+ = M_{p}^{2}-{{\left( \frac{\sqrt{s}}{2}+{{q}^{0}} \right)}^{2}}+{{q}^{2}}=0 \\ 
& z_1^- = M_{p}^{2}-{{\left( \frac{\sqrt{s}}{2}-{{q}^{0}} \right)}^{2}}+{{q}^{2}}=0 \\ 
\end{split} \right.\
\end{equation}

It is easy to verify that ${{q}^{0}}=0,\;q=P$ satisfies the equation system \eqref{systema_rivnan}. This solution describes the $2$-dimensional sphere and thus provides some subset of the $4$-dimensional integration domain where the absolute value of the first two denominators in \eqref{tensor_in_sphere} has a single minimum.

Next, let us apply Laplace's method to the first pair of denominators
\begin{equation}\label{exponenta_vid_log}
\begin{split}
&\frac{1}{\left(z_1^+ - i\varepsilon\right)\left(z_1^- - i\varepsilon\right)}
=
\\
& = 
\exp{\left[-\ln{\left(z_1^+-i\varepsilon\right)} -\ln{\left(z_1^--i\varepsilon\right)} \right]} = 
\\ 
& =
-\frac{1}{\varepsilon^2}
\exp{
\left[
- \ln{\left(1- \frac{z_1^+}{i\varepsilon}\right)} 
- \ln{\left(1 - \frac{z_1^-}{i\varepsilon}\right)} 
\right].
}
\end{split}
\end{equation}

Considering the second order Taylor approximation of the exponent in \eqref{exponenta_vid_log} and changing the variable $q = P + x$ in \eqref{zpm_in_sphere}, we obtain
\begin{equation}\label{exponenta_vid_log1}
\begin{split}
\frac{1}{\left(z_1^+ - i\varepsilon\right)\left(z_1^- - i\varepsilon\right)}
 \approx
 -\frac{1}{\varepsilon^2}
\exp \left[
\frac{s{{\left( {{q}^{0}} \right)}^{2}}+4{{P}^{2}}{{x}^{2}}}{-{{\varepsilon }^{2}}} \right.
\\
\hphantom{\hspace{2cm}}
\left.
+\frac{-2{{\left( {{q}^{0}} \right)}^{2}}+4Px+2{{x}^{2}}}{i\varepsilon } 
\right].
\end{split}
\end{equation}
The convenience of variable $x = q - P$ is that together with $q^0$ they are the offset from the point $(q^0 = 0,\; q = P)$ where the real parts of denominators in the left-hand side of \eqref{exponenta_vid_log1} take minimum value. Consequently, the absolute value of exponent in \eqref{exponenta_vid_log1} has maximum at $\left(q^0 = 0,\; x = 0 \right)$. 

Changing the variable $q \to x = q - P$ in the integral \eqref{tensor_in_sphere} and substituting \eqref{exponenta_vid_log1}, we get the approximation for the integral \eqref{tensor_in_sphere}
\begin{widetext}
\begin{equation}\label{tensor_for_z12} 
\begin{split}
 {{t}_{ab}} \approx &
-\frac{1}{\varepsilon^2} 
\int\limits_{-\infty }^{\infty }{d{{q}^{0}}}
\int\limits_{-P}^{\infty }{{{\left(P + x\right)}^{2}}dx}
\int\limits_{0}^{\pi }{\sin \left( \theta  \right)}d\theta
\int\limits_{0}^{2\pi }{d\varphi }
\left[ 
\vphantom{\frac{1}{\left(z_3^{-} - i\varepsilon\right)\left(z_3^{+} - i\varepsilon\right)}}
{{f}_{ab}}\left( {{q}^{0}},x,\theta ,\varphi  \right) \right.
 \\
& \left.
\times
\exp{\left(
\frac{-2{{\left( {{q}^{0}} \right)}^{2}}+4Px+2{{x}^{2}}}{i\varepsilon }+\frac{s{{\left( {{q}^{0}} \right)}^{2}}+4{{P}^{2}}{{x}^{2}}}{-{{\varepsilon }^{2}}}
\right)}
\frac{1}{\left(z_3^{-} - i\varepsilon\right)\left(z_3^{+} - i\varepsilon\right)}
\right],
\end{split}
\end{equation}
\end{widetext}
\noindent
where the same variable change $q = P + x$ is also performed for $f_{ab}, \; z_3^+$ and $z_3^-$.


\begin{figure*}
	\centering
	\includegraphics[width=0.44\linewidth]{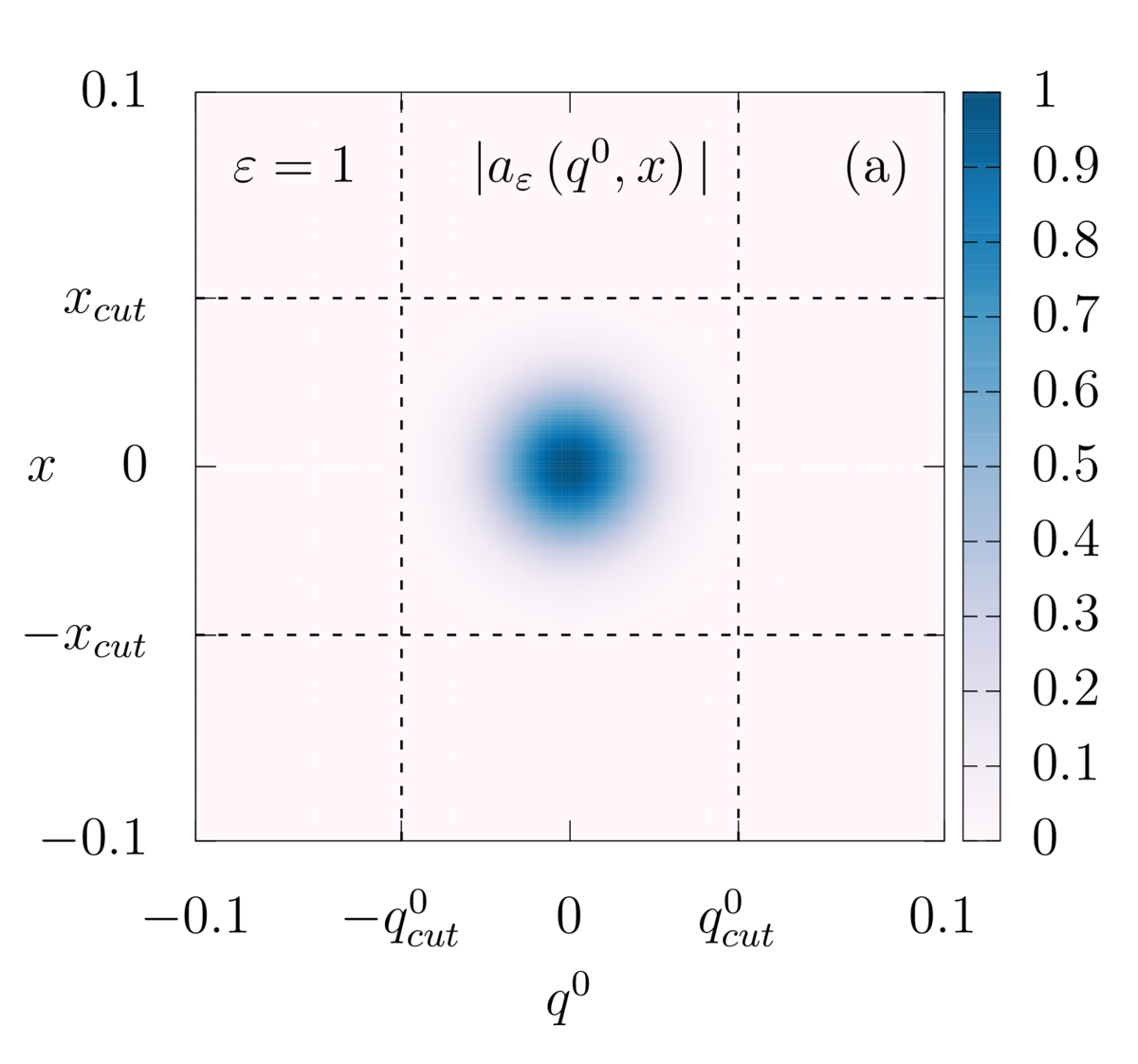}
	\includegraphics[width=0.44\linewidth]{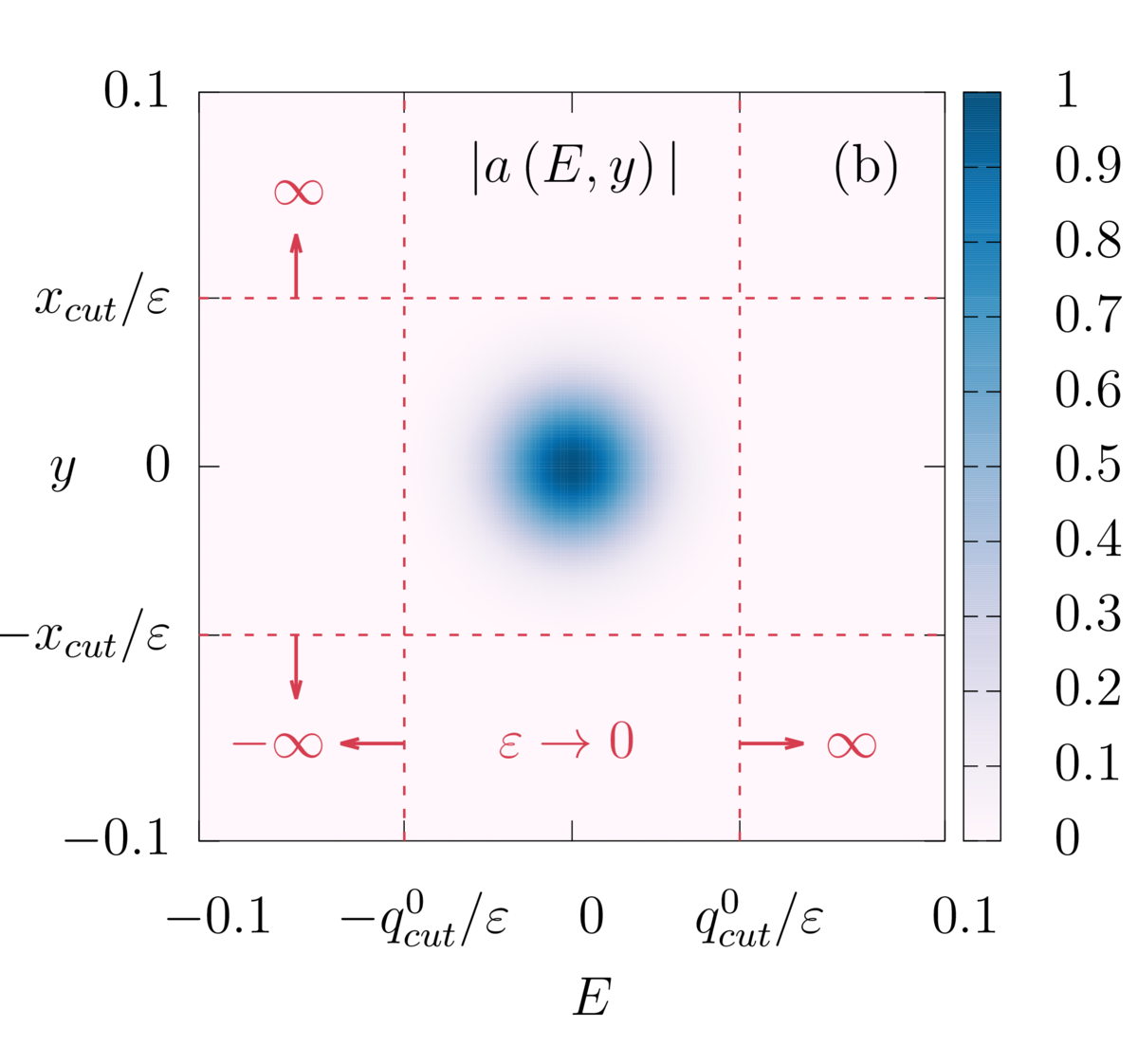}
	\caption{Colormap plot of $\vert a_\varepsilon\left(q^0, x\right) \vert $ in $q^0x$ plane (a), and 
	$\vert a\left(E, y\right) \vert$ in $Ey$ plane (b) for $\varepsilon = 1$. The values of parameters: $\sqrt{s} = 44.6$ GeV, $M_p = 0.938$ GeV.}
	\label{fig:a_abs}
\end{figure*}

We next turn to analyzing the integrand in \eqref{tensor_for_z12}. We denote it shortly by $a_\varepsilon\left(q^0, x, \theta, \phi\right)$. The exponent \eqref{exponenta_vid_log1} in the integrand depends only on the integration variables $q^0$ and $x$. The absolute value of this exponent has maximum at $\left(q^0 = 0, \; x = 0\right)$, and the \textit{width} of its peak tends to zero as $\varepsilon \to 0$. Since the other part of the integrand \eqref{tensor_for_z12} containing the second pair of denominators does not have singularities at $\left(q^0 = 0, \; x = 0\right)$, we can use such the exponential suppression of the integrand to cut the integration domain 
\begin{equation}\label{domain_cut} 
\int^\infty_{-\infty}dq^0\int^\infty_{-P}dx  \to \int^{q^0_{cut}}_{-q^0_{cut}}dq^0\int^{x_{cut}}_{-x_{cut}}dx,
\end{equation}
\noindent
where $q^0_{cut}  > 0$ and $0 < x_{cut} < P$ .

This idea is demonstrated in \fref{fig:a_abs}a. First, we found that the function $a_\varepsilon\left(q^0 = 0, x = 0, \theta, \phi\right)$ has maximum at $(\theta = 0, \phi = 0)$. Then we considered a restriction of $a_\varepsilon\left(q^0, x, \theta, \phi \right)$ to $a_\varepsilon\left(q^0, x, \theta = 0, \phi = 0\right)$ which we denoted by $a_\varepsilon\left(q^0, x\right)$ and plotted its absolute value in the $q^0x$ plane.

In this way we consider the contribution to the original integral \eqref{tensor_in_sphere} supplied by the first pair of denominators $z_1$ and $z_2$. Note that the actual value of this contribution does not depend on the selection of $q^0_{cut}$ and $x_{cut}$, which becomes clear as soon as we make the following change of variables
\begin{equation}\label{zamina1}
\begin{aligned}[c]
x=\varepsilon y
\end{aligned}
\qquad \qquad
\begin{aligned}[c]
{{q}^{0}}=\varepsilon E.
\end{aligned}
\end{equation}
\noindent
It makes the integration \eqref{domain_cut} to take the form
\begin{equation}\label{domain_cut_yE} 
\int^{q^0_{cut}}_{-q^0_{cut}}dq^0
\int^{x_{cut}}_{-x_{cut}}dx  
\to
\varepsilon^2
\int^{q^0_{cut}/\varepsilon}_{-q^0_{cut}/\varepsilon}dE
\int^{x_{cut}/\varepsilon}_{-x_{cut}/\varepsilon}dy
\end{equation}
\noindent
and also cancels all the $\varepsilon$ in denominators in \eqref{tensor_for_z12}. Since there is no $\varepsilon$ in the denominator now, we can finally pass to the limit when $\varepsilon \to 0$. At this point it is clear that the resulting integration limits will not depend on the particular selection of $q^0_{cut}$ and $x_{cut}$ because both $q^0_{cut}/\varepsilon$ and $x_{cut}/\varepsilon$ tend to infinity as $\varepsilon \to 0$ (see \fref{fig:a_abs}b), while the \textit{width} of the peak of 
$\vert a\left(E, y\right) \vert$ remains unchanged.

Now \eqref{tensor_for_z12} can be rewritten as
\begin{equation}\label{tensor7}
\begin{split}
 {{t}_{ab}}
\approx
&-{{P}^{2}}
\int\limits_{-\infty }^{\infty }{dE}
\int\limits_{-\infty }^{\infty }{dy}
\int\limits_{0}^{\pi }{\sin \left( \theta  \right)}d\theta 
\int\limits_{0}^{2\pi }{d\varphi }
\left[
\vphantom{
\frac{1}{\left(z_3^{-} - i\varepsilon\right)\left(z_3^{+} - i\varepsilon\right)}
}
{{f}_{ab}}\left(\theta ,\varphi  \right) 
\right. 
\\ 
&\left.
\times 
\exp \left( -4iPy-s{{E}^{2}}-4{{P}^{2}}{{y}^{2}} \right) 
\frac{1}{\left(z_3^{-} z_3^{+} \right)}
\right],
\end{split}
\end{equation}
\noindent
where $f_{ab}\left(\theta, \varphi\right)$ now denotes $f_{ab}\left(q^0 = 0, x = 0, \theta, \varphi\right)$. The integrals over $E$ and $y$ in \eqref{tensor7} are now reduced to Poisson integrals
\begin{equation}\label{rozrahunok_puason}
\begin{split}
& \int\limits_{-\infty }^{\infty }{dE}\exp \left( -s{{E}^{2}} \right)=\frac{\sqrt{\pi }}{\sqrt{s}}, \\
& \int\limits_{-\infty }^{\infty }{dy}\exp \left( -4{{P}^{2}}{{y}^{2}}-4iPy \right)= \frac{\sqrt{\pi }}{2 e P},
\end{split}
\end{equation}
\noindent
where $e$ is the Euler's number (i.e. the base of the natural logarithm).

Combining \eqref{tensor7} with \eqref{rozrahunok_puason} and substituting expressions for $z_3^\pm$
we obtain
\begin{widetext}
\begin{equation}\label{tensor8}
\begin{split}
 {{t}_{ab}}\approx & -\frac{\pi P}{2e\sqrt{s}}\int\limits_{0}^{\pi }{\sin \left( \theta  \right)d\theta }\int\limits_{0}^{2\pi }{d\varphi }
\left[
\vphantom{\frac{a^2}{b^2}}
 {{f}_{ab}}\left( \theta ,\varphi  \right) \right.  \\ 
& \left. \times \frac{1}{{{\left( M_{G}^{2}+2{{P}^{2}}-P\cos \left( \theta  \right)\sqrt{4{{P}^{2}}-\left| t \right|} \right)}^{2}}-{{P}^{2}}\left| t \right|{{\sin }^{2}}\left( \theta  \right){{\cos }^{2}}\left( \varphi  \right)}
\right].
\end{split}
\end{equation}
\end{widetext}

Finally, there remains only two-dimensional integral over $\theta$ and $\varphi$ for each component $t_{ab}$. Moreover, we can calculate the integral over $\varphi$ for each pair of indices $(a, b)$ analytically (these calculations are quite long but yet straightforward). So we end up with one-dimensional integral over $\theta$ which can be calculated numerically.

\subsection{The second pair of denominators}
Let us now consider the region of the integration domain in \eqref{tensor_ab_short}, where the divergence of the integrand  arises due to the second pair of denominators
\begin{equation}
\begin{split}
&\frac{1}{M_{G}^{2}-{{\left( {{k}^{0}} \right)}^{2}}+{{{\vec{k}}}^{2}}-i\varepsilon }
\\
\times
&\frac{1}{M_{G}^{2}-{{\left( {{k}^{0}} \right)}^{2}}+{{\left( \left( {{{\vec{P}}}_{1}}-{{{\vec{P}}}_{3}} \right)-\vec{k} \right)}^{2}}-i\varepsilon },
\end{split}
\label{second_denom_pair}
\end{equation}
\noindent
If we let both the energy component ${{k}^{0}}$ of four-momentum $k$ and the length of its spatial part $| {\vec{k}} |$ tend to infinity simultaneously, the difference of their squares ${{\left( {{k}^{0}} \right)}^{2}}-{{| {\vec{k}} |}^{2}}$ may still be finite. In contrast, the sum ${{\left( {{k}^{0}} \right)}^{2}} + {{| {\vec{k}} |}^{2}}$ takes infinite values even when either ${{k}^{0}}$ or $| {\vec{k}}|$ tends to infinity. Note that the first pair of denominators contains ${{k}^{0}}$ in first power, which provides the decreasing of the integrand as ${{k}^{0}}$ tends to infinity. At the same time, in the center of mass system the energy component of ${{P}_{1}}-{{P}_{3}}$ is equal to zero, and the denominators \eqref{second_denom_pair} do not contain the first power of ${{k}^{0}}$, but only the difference ${{\left( {{k}^{0}} \right)}^{2}}-{{| {\vec{k}} |}^{2}}$ which, as we have already mentioned, may take finite values as ${{k}^{0}} \to \infty$ and $| {\vec{k}} | \to \infty$. In this case the whole integrand goes to zero due to the first pair of denominators containing the first power of ${{k}^{0}}$. 

If we consider the component $t_{00}$ of the tensor \eqref{tensor_ab_short}, the numerator ${{f}_{ab}}\left( {{k}^{0}},\vec{k} \right)=s-{{\left( {{k}^{0}} \right)}^{2}}$, where $s$ is the Mandelstam invariant, tends to infinity as the denominator of the integrand. In this case the whole integral diverges. However, this divergence of \eqref{tensor_ab_short} is only
delusive. Let us show that.

As mentioned above, if the expression contained the sum instead of the difference ${{\left( {{q}^{0}} \right)}^{2}}-{{\left| {\vec{q}} \right|}^{2}}$, there would be no problem with the divergence at all. It is possible to turn this difference into a sum if we succeeded by transforming the integration contour in the complex plane from the real axis to the imaginary one, which can be done by the Wick rotation. The possibility of this transformation depends on the location of the integrand poles. Les us take a closer look at them.

First, we rewrite the expression \eqref{tensor_t_through_q} in the form
\begin{widetext}
\begin{equation}\label{tensor10}
\begin{split}
{{t}_{ab}}= & \int\limits_{-\infty }^{\infty }{d{{q}_{0}}}\int\limits_{-\infty }^{\infty }{d\vec{q}}
\left[
{{f}_{ab}}\left( {{q}_{0}},\vec{q} \right) 
\frac{1}{{{q}^{0}}-\left( -\frac{\sqrt{s}}{2}+ \eta \right)}
 \frac{1}{{{q}^{0}}-\left( -\frac{\sqrt{s}}{2}- \eta \right)}
 \right.
 \\
 & \times 
 \left.
 \frac{1}{{{q}^{0}}-\left( \frac{\sqrt{s}}{2}+ \eta \right)}  
 \frac{1}{{{q}^{0}}-\left( \frac{\sqrt{s}}{2}- \eta \right)}
 \frac{1}{\left(z_3 - i\varepsilon\right)\left(z_4 - i\varepsilon\right)} 
 \right] ,
\end{split}
\end{equation}
\end{widetext}
\noindent
where $\eta = \sqrt{M_{p}^{2}+{{\left( {{q}^{1}} \right)}^{2}}+{{\left( {{q}^{2}} \right)}^{2}}+{{\left( \frac{\sqrt{4{{P}^{2}}-\left| t \right|}}{2}+{{q}^{3}} \right)}^{2}}-i\varepsilon }$. 

Let us analyze the placement of quantities $\eta$ and $-\eta$ on the complex plane. It depends on the variables $ {{q}^{1}},{{q}^{2}},{{q}^{3}}$, so it varies across different subsets of the ${{q}^{1}},{{q}^{2}},{{q}^{3}}$ integration domain. Thus the major question here is which of the two conditions
\begin{equation}
\begin{aligned}[c]
\eta {\vert_{\varepsilon = 0}}  > \frac{\sqrt{s}}{2}
\end{aligned}
\qquad
\begin{aligned}[c]
\text{or}
\end{aligned}
\qquad 
\begin{aligned}[c]
\eta {\vert_{\varepsilon = 0}} <\frac{\sqrt{s}}{2}
\end{aligned}
\label{quadrants_ineq}
\end{equation}
\noindent
is met for the particular subset. Whether the first or the second condition is met, determines the distribution of the poles among the quadrants of the complex plane.

In case the second inequality \eqref{quadrants_ineq} holds, each quadrant of the complex plane contains a pole of the integrand \eqref{tensor10}. It means that one cannot do Wick rotation, because one would cross the poles during rotation. 

In the case of the first inequality, the poles are located in the second and fourth quadrants, so it is possible to rotate the integration path in the first and third quadrants. Thus the integration over ${{q}^{0}}$ can be transformed into the integration along the imaginary axis (Wick rotation), and the expression ${{\left( {{q}^{0}} \right)}^{2}}-{{\left| {\vec{q}} \right|}^{2}}$ transforms into $\left( -{{\left( {{q}^{0}} \right)}^{2}}-{{\left| {\vec{q}} \right|}^{2}} \right)$ as the integration variable changes ${{q}^{0}} \rightarrow i{{q}^{0}}$. If one of these components tends to infinity, their sum also tends to infinity, which solves the problem. The integrand tends to zero as ${{\left( {{\left( {{q}^{0}} \right)}^{2}}+{{\left| {\vec{q}} \right|}^{2}} \right)}^{-4}}$, which guarantees the convergence of the integral.

The possibility to apply Wick rotation arises in the region of high values of $ \left|\vec{q}\right|$, i.e. in that part of the integration domain where the integral can diverge. Therefore, selecting this region and applying corresponding Wick rotation, we obtain the convergent integral. The rest of the integration domain is finite and does not impact the convergence of the integral, since there are no singularities of the integrand in this region at $ \varepsilon \neq 0$. This was actually the reason for calling the divergence of the integral ``delusive'' earlier in this section.

\subsection{Calculation of the second contribution with the Wick rotation}
Let us implement the idea explained in the previous section. We split the integral \eqref{tensor_t_through_q} into two integrals. The first integral over the finite region, where the Wick rotation cannot be applied, we denote as
\begin{equation}\label{tensordovika}
\begin{split}
& {{t}_{ab}^{<}}=\int\limits_{-\infty }^{\infty }{d{{q}^{0}}}\int\limits_{\left| {\vec{q}} \right|<cP}{d\vec{q}}{{f}_{ab}}\left( {{q}^{0}},\vec{q} \right)
\prod_{j=1}^4\frac{1}{\left(z_j - i\varepsilon\right)},
\end{split}
\end{equation} 
\noindent
where $c > 1$. The value of $c$ has no impact on the result of integration. However, if $c=1$ and $\left| {\vec{q}} \right|<P$, then the poles of integrand \eqref{tensor10} lie on the integration path and the integral diverges, which imposes the lower bound on $c$.

The second contribution to the tensor ${{t}_{ab}}$, i.e. the integral over the region where the Wick rotation can be applied, we denote as
\begin{equation}\label{tensorposlevika}
\begin{split}
& {{t}_{ab}^{>}}=\int\limits_{-\infty }^{\infty }{d{{q}^{0}}}\int\limits_{\left| {\vec{q}} \right|>cP}{d\vec{q}}{{f}_{ab}}\left( {{q}^{0}},\vec{q} \right)
\prod_{j=1}^4\frac{1}{\left(z_j - i\varepsilon\right)}. 
\end{split}
\end{equation}
\noindent
Here we also use the notations \eqref{zpm} for denominators with $\vec{P}_1$ and $\vec{P}_3$ expressed through $t$ \eqref{cherez_t}
\begin{equation}\label{zpm2}
\begin{split}
 z_1^{\pm}=M_{p}^{2}-{{\left( \frac{\sqrt{s}}{2}\pm{{q}^{0}} \right)}^{2}}+{{\left( {{q}^{1}} \right)}^{2}}+{{\left( {{q}^{2}} \right)}^{2}}
\\ 
\hspace{2cm}
+{{\left( \frac{\sqrt{4P^2 - \vert t \vert}}{2}+{{q}^{3}} \right)}^{2}},
\\
z_3^{\pm} =  M_{G}^{2}-{{\left( {{q}^{0}} \right)}^{2}}+{{\left( \frac{\sqrt{|t|}}{2} \pm {{q}^{1}} \right)}^{2}}
\hphantom{11+1}
\\
\hspace{3cm}
+{{\left( {{q}^{2}} \right)}^{2}}+{{\left( {{q}^{3}} \right)}^{2}} .
\end{split}
\end{equation} 

Next we consider the system of equations for the second pair of denominators similar to what we did for the first pair of denominators. First, we set the real parts equal to zero and then find the second conditional maxima
\begin{equation}\label{vtoroymax}
\left\{ \begin{split}
& z_3^- = M_{G}^{2}-{{\left( {{q}^{0}} \right)}^{2}}+{{\left( \frac{\sqrt{\left| t \right|}}{2}-{{q}^{1}} \right)}^{2}}+{{\left( {{q}^{2}} \right)}^{2}}+{{\left( {{q}^{3}} \right)}^{2}}=0 \\ 
& z_3^+ = M_{G}^{2}-{{\left( {{q}^{0}} \right)}^{2}}+{{\left( \frac{\sqrt{\left| t \right|}}{2}+{{q}^{1}} \right)}^{2}}+{{\left( {{q}^{2}} \right)}^{2}}+{{\left( {{q}^{3}} \right)}^{2}}=0. \\ 
\end{split}\right.\
\end{equation} 
\noindent
Let $\chi = \sqrt{M_{G}^{2}+\frac{\left| t \right|}{4}+{{\left( {{q}^{2}} \right)}^{2}}+{{\left( {{q}^{3}} \right)}^{2}}}$. It is easily seen that ${{q}^{1}}=0$, ${{q}^{0}}=\pm \chi$ is the solution of \eqref{vtoroymax}.  Note that ${{q}^{0}}=\pm \chi$ can take both positive and negative values, but we consider only positive solutions, since the integrand is an even function of ${{q}^{0}}$.  We can now proceed the calculation of the integral \eqref{tensordovika} analogously to what we did in Section \ref{first_pair_section}. First, we represent the second pair of denominators in exponential form
\begin{equation}\label{z34_exp_form}
\begin{split}
 \frac{1}{z_3^\pm - i\varepsilon}
 = &
 \frac{1}{
    \sqrt{
        \left(z_3^\pm\right)^2 + \varepsilon^2 
    } 
 }
 \\
 & \times
 \exp{
   \left(
   i 
   \arccos{
      \frac
      {z_3^\pm}
      {\sqrt{\left(z_3^\pm\right)^2 + \varepsilon^2 }}
   }
   \right) 
 }.
\end{split}
\end{equation}
\noindent
In order to cancel the $\varepsilon$ in denominators in \eqref{z34_exp_form}, we make the following variable change (analogously to \eqref{zamina1})
\begin{equation}\label{zamina_2}
\begin{aligned}[c]
{{q}^{0}}= \chi + \varepsilon E
\end{aligned}
\qquad \qquad
\begin{aligned}[c]
{{q}^{1}}=\varepsilon x,
\end{aligned}
\end{equation}
\noindent
so after passing to the limit when $\varepsilon \to 0$, the right-hand side of the expression \eqref{z34_exp_form} takes the form
\begin{equation}\label{z3pm_t1}
\begin{split}
\frac{1}{z_3^\pm - i\varepsilon}\bigg|_{\varepsilon = 0}
= &
\frac{1}{\sqrt{\omega_\pm^2 + 1}}
\\
& \times
\exp{
\left(
i \arccos {
\frac{\omega_\pm}{\sqrt{\omega_\pm^2 + 1}}
}
\right)
},
\end{split}
\end{equation}
\noindent
where $\omega_\pm = -2E\chi \pm \sqrt{|t|}x$.

Then we can rewrite the integral \eqref{tensordovika} as
\begin{widetext}
\begin{equation}\label{tensordovika2}
\begin{split}
 {{t}_{ab}^{<}}=
 2 & \int\limits_{-\infty }^{\infty }{dE}
\int\limits_{-\infty }^{\infty }{dx}
\int\limits_{D^<}{d{{q}^{2}}d{{q}^{3}}}
\left[ \vphantom{\frac{1^1}{1^1}}
{{f}_{ab}}\left( {{q}^{2}},{{q}^{3}} \right)
\frac{1}{\left(z_1^+ z_1^-\right)}
\frac{1}{\sqrt{\omega_-^2 + 1}}
\right.
 \\ 
& \times  \left.
\frac{1}{\sqrt{ \omega_+^2 + 1}}
\exp{ 
\left( i \arccos {
\frac{\omega_-}{\sqrt{\omega_-^2 + 1}}
}
+
i \arccos {
\frac{\omega_+}{\sqrt{\omega_+^2 + 1}}
}
\right)
}
\right],
\end{split}
\end{equation}
\end{widetext}
\noindent
where $D^< = \lbrace{ \left(q^2, q^3\right) \in R^2 \; \vert \;  {{\left( {{q}^{2}} \right)}^{2}}+{{\left( {{q}^{3}} \right)}^{2}}<{c}^{2}{{P}^{2}}  \rbrace}$ and ${{f}_{ab}}\left( {{q}^{2}},{{q}^{3}} \right)$ denotes ${{f}_{ab}}\left(q^0 = \chi, q^1 = 0, {{q}^{2}},{{q}^{3}} \right)$.

According to Laplace's method \cite{DeBruijn:225131}, we transform the square roots in the denominators in \eqref{tensordovika2} containing $\omega_\pm$ as
\begin{equation}\label{omega_laplace1}
\begin{split}
\frac{1}{\sqrt{ \omega_+^2 + 1}\sqrt{ \omega_-^2 + 1}} = 
\exp
&
\left[
-\frac{1}{2}\log
\left(
\omega_+^2 + 1
\right)
\right.
\\
&
\left.
-\frac{1}{2}\log
\left(
\omega_-^2 + 1
\right)
\right],
\end{split}
\end{equation}
\noindent
and take the second order Taylor approximation of the exponent at $E = 0, \; x = 0$
\begin{equation}\label{omega_laplace2}
\frac{1}{\sqrt{ \omega_+^2 + 1}\sqrt{ \omega_-^2 + 1}} 
\approx 
\exp \left( -4{{E}^{2}} \chi ^ 2 -\left| t \right|{{x}^{2}} \right).
\end{equation}
\noindent
Similarly, for the other two terms in these denominators
\begin{equation}\label{exp_omega_laplace1}
\begin{split}
\exp{ 
\left( i \arccos {
\frac{\omega_-}{\sqrt{\omega_-^2 + 1}}
}
+
i \arccos {
\frac{\omega_+}{\sqrt{\omega_+^2 + 1}}
}
\right)
}
\approx
\\ 
\approx
-
\exp{
\left(
-4i
E\chi
\right).
}
\end{split}
\end{equation}
Substituting \eqref{omega_laplace2} and \eqref{exp_omega_laplace1} into \eqref{tensordovika2}, the integration with respect to $E$ and $x$ is now reduced to the Poisson integrals
\begin{equation}\label{puason1}
\begin{split}
&\int\limits_{-\infty }^{\infty }{dx}\exp \left( -\left| t \right|{{x}^{2}} \right)=\frac{\sqrt{\pi }}{\sqrt{\left| t \right|}}
\\
& \int\limits_{-\infty }^{\infty }{dE}\exp \left( -4{{E}^{2}}\chi^2 + 4iE\chi \right) = 
\frac{ \sqrt{\pi } } { 2 e \chi }.
\end{split}
\end{equation}
After these transformations we can rewrite the expression \eqref{tensordovika2} as
\begin{equation}\label{tensordovika3}
\begin{split}
 {{t}_{ab}^<} 
\approx &
-\frac{\pi }{2e\sqrt{\left| t \right|}}
\int\limits_{D^<}{d{{q}^{2}}d{{q}^{3}}}
\left[
\frac{{{f}_{ab}}\left( {{q}^{2}},{{q}^{3}} \right)}{\chi}
\right.
\\
& \times
\left.
 \frac{1}{{{\left( M_{G}^{2}+\frac{\left| t \right|}{2}-{{q}^{3}}\sqrt{4{{P}^{2}}-\left| t \right|} \right)}^{2}}-s\chi^2}
 \right].
\end{split}
\end{equation}
It is convenient to use the polar coordinates for further calculations
\begin{equation}\label{polyarnye}
\begin{split}
& {{q}^{2}}=q\cos \left( \alpha  \right), \\ 
& {{q}^{3}}=q\sin \left( \alpha  \right). \\ 
\end{split}
\end{equation}
Then the expression \eqref{tensordovika3} takes the form 
\begin{widetext}
\begin{equation}\label{tensordovika4}
\begin{split}
& {{t}_{ab}^{<}}=-\frac{\pi }{2e\sqrt{\left| t \right|}}\int\limits_{0}^{cP}{qdq\frac{1}{\sqrt{M_{G}^{2}+\frac{\left| t \right|}{4}+{{q}^{2}}}}}\times  \\ 
& \times \int\limits_{0}^{2\pi }{d\alpha }\frac{{{f}_{ab}\left( q,\alpha  \right)}}{{{\left( M_{G}^{2}+\frac{\left| t \right|}{2}-q\sin \left( \alpha  \right)\sqrt{4{{P}^{2}}-\left| t \right|} \right)}^{2}}-s\left( M_{G}^{2}+\frac{\left| t \right|}{4}+{{q}^{2}} \right)}.
\end{split}
\end{equation}
\end{widetext}

As a result, we obtain the two-dimensional integral with respect to the variables $q$ and $\alpha$. This integral can be calculated numerically and allows us to calculate each component of the tensor ${{t}_{ab}}$ separately.

Let us get back to the expression \eqref{tensorposlevika}. Applying Wick rotation, we change the variable ${{q}^{0}}\to i{{q}^{0}}$ in the integral \eqref{tensorposlevika}, so it can be rewritten as
\begin{equation}\label{tensorposlevika1}
\begin{split}
 {{t}_{ab}^{>}}=
i
\int\limits_{-\infty }^{\infty }{d{{q}^{0}}}
\int\limits_{-\infty }^{\infty }{d{{q}^{2}}}
\int\limits_{D^>}{d{{q}^{1}}}d{{q}^{3}}
\left[
\vphantom{\frac{1^1}{1^1}}
{{f}_{ab}}\left( {{q}^{0}},{{q}^{1}},{{q}^{2}},{{q}^{3}} \right)
\right.
\\
\left.
\times 
\frac{1}{\left(z_1 - i\varepsilon\right)}
\frac{1}{\left(z_2 - i\varepsilon\right)}
\frac{1}{\left(z_3 - i\varepsilon\right)}
\frac{1}{\left(z_4 - i\varepsilon\right)}
\right]
,
\end{split}
\end{equation}
where $D^> = \lbrace{ \left(q^1, q^3\right) \in R^2 \; \vert \;  {{\left( {{q}^{1}} \right)}^{2}}+{{\left( {{q}^{3}} \right)}^{2}}>{c}^{2}{{P}^{2}}  \rbrace}$.

 Note that $\varepsilon $ in denominators can be set to zero now, because the integration is performed along the imaginary axis and we obtain the convergent integral. Rewriting the denominators in \eqref{tensorposlevika1}, we obtain
\begin{equation}\label{tensorposlevika2}
\begin{split}
 {{t}_{ab}^{>}}=
i
& \int\limits_{-\infty }^{\infty }{d{{q}^{0}}}
\int\limits_{-\infty }^{\infty }{d{{q}^{2}}}
\int\limits_{D^>}{d{{q}^{1}}}d{{q}^{3}}
\left[
\vphantom{\frac{1^1}{1^1}}
{{f}_{ab}}\left( {{q}^{0}},{{q}^{1}},{{q}^{2}},{{q}^{3}} \right)
\right.
\\ 
 & \times
 \frac{1}{{{\left(M_{p}^{2} + \xi^2 + {{\left(\frac{\sqrt{4P^2 - |t|}}{2} + {{q}^{3}} \right)}^{2}}-\frac{s}{4} \right)}^{2}}+\left(q^0\right)^2s} \\\ 
&  \left.
\times 
\frac{1}{{{\left( M_{G}^{2}+ \frac{|t|}{4}+ \xi^2 +{{\left( {{q}^{3}} \right)}^{2}} \right)}^{2}}-{{\left( {{q}^{1}} \right)}^{2}} |t|}
\right]
,
\end{split}
\end{equation}
\noindent
where $\xi^2 = {{\left( {{q}^{0}} \right)}^{2}} + {{\left( {{q}^{1}} \right)}^{2}} + {{\left( {{q}^{2}} \right)}^{2}}$.

It can be seen that the product of two fractions in \eqref{tensorposlevika2} is even function with regard to $q^0$ and $q^2$ and 
has the maximum at ${{q}^{0}}=0, {{q}^{2}}=0$. Increasing the absolute values of ${{q}^{0}}$ and ${{q}^{2}}$ results in increasing the denominators, i.e. in decreasing the whole fraction. Therefore we return back to \eqref{tensorposlevika1} and separate the integration with respect to ${{q}^{0}}, {{q}^{2}}$ and ${{q}^{1}}, {{q}^{3}}$. For the convenience and simplicity of the further calculations, we introduce the following polar coordinates
\begin{equation}\label{polyarnye1}
\begin{aligned}[c]
{{q}^{1}}=q\cos \left( \alpha  \right)
\end{aligned}
\quad \quad
\begin{aligned}[c]
{{q}^{3}}=q\sin \left( \alpha  \right).
\end{aligned}
\end{equation}
\noindent
The expression \eqref{tensorposlevika1} then can be rewritten as
\begin{equation}\label{tensorposlevika3}
\begin{split}
 {{t}_{ab}^{>}}=i\int\limits_{cP}^{\infty }{qdq}\int\limits_{0}^{2\pi }{d\alpha }\int\limits_{-\infty }^{\infty }{d{{q}^{0}}}
 & \int\limits_{-\infty }^{\infty }{d{{q}^{2}}}
\left[ \vphantom{\frac{1}{1}}
{{f}_{ab}}\left( q,\alpha ,{{q}^{0}},{{q}^{2}} \right) 
\right.
\\
& \times
\left.
\frac{1}{z_1^+z_1^-}
\frac{1}{z_3^+z_3^-}
\right],
\end{split}
\end{equation}
\noindent
where
\begin{equation}\label{z1z3_wick}
\begin{split}
z_1^\pm = M_{p}^{2}+ 
& 
{{\left( {{q}^{0}} \mp i\frac{\sqrt{s}}{2} \right)}^{2}}+{{q}^{2}}{{\cos }^{2}}\left( \alpha  \right)+{{\left( {{q}^{2}} \right)}^{2}}
\\ 
& 
\hphantom{{{\left( 1 \right)}^{2}}}
+{{\left( \frac{\sqrt{4P^2 - |t|}}{2}+q\sin \left( \alpha  \right) \right)}^{2}},
\\
 z_3^\pm = M_{G}^{2}+ 
 & 
{{\left( {{q}^{0}} \right)}^{2}}+{{\left( \frac{\sqrt{|t|}}{2}\mp q\cos \left( \alpha  \right) \right)}^{2}}
\\
& 
\hphantom{{{\left({{q}^{0}}  {{q}^{0}} \right)}^{2}}}
+{{\left( {{q}^{2}} \right)}^{2}}+{{q}^{2}}{{\sin }^{2}}\left( \alpha  \right).
\end{split}
\end{equation}

The expression \eqref{tensorposlevika3} contains the integration with respect to $q$ from $cP$ to $+\infty$. We can change the variable $q \to x = 1 / q$ in order to make the integration limits finite. At this point the integrand in \eqref{tensorposlevika3} becomes large enough if we substitute the expressions for \eqref{z1z3_wick} for $z_1^\pm$ and $z_3^\pm$ which depend on every integration variable $x, \alpha, q^0, q^2$. However, introducing the following notations
\begin{equation}\label{omega_xalpha}
\begin{split}
& {{w}_{1}}\left( x,\alpha  \right)=1-{{x}^{2}}\frac{\left| t \right|}{4}+x\sqrt{4{{P}^{2}}-\left| t \right|}\sin \left( \alpha  \right) \\ 
& {{w}_{2}}\left( x,\alpha  \right)={{x}^{2}}M_{G}^{2}+{{x}^{2}}\frac{\left| t \right|}{4}-x\sqrt{\left| t \right|}\cos \left( \alpha  \right)+1 \\ 
& {{w}_{3}}\left( x,\alpha  \right)={{x}^{2}}M_{G}^{2}+{{x}^{2}}\frac{\left| t \right|}{4}+x\sqrt{\left| t \right|}\cos \left( \alpha  \right)+1,
\end{split}
\end{equation}
\noindent
and changing the integration variables again $q^0 \to E = x{{q}^{0}}$ and $q^2 \to y = x{{q}^{2}}$, we can finally rewrite \eqref{tensorposlevika3} using notations \eqref{omega_xalpha}
\begin{equation}\label{tensorposlevika6}
\begin{split}
 {{t}_{ab}^{>}}=&\hphantom{a}i
\int\limits_{0}^{\frac{1}{cP}}{xdx}
\int\limits_{0}^{2\pi }{d\alpha }
\int\limits_{-\infty }^{\infty }{dE}
\int\limits_{-\infty }^{\infty }{dy}
\left[\vphantom{\frac{1^1}{1^1}}
{{f}_{ab}}\left( x,E,\alpha ,y \right)
\right.
\\
& \times
\frac{1}{{{\left( {{w}_{1}}\left( x,\alpha  \right) \right)}^{2}}{{w}_{2}}\left( x,\alpha  \right){{w}_{3}}\left( x,\alpha  \right)}
\\ 
& \times
\left. 
\frac{1}{1+{{W}_{1}}\left( x,\alpha  \right)\left( {{E}^{2}}+{{y}^{2}} \right)+{{W}_{2}}\left( x,\alpha  \right){{E}^{2}}}
\right],
\end{split}
\end{equation}
\noindent
where 
\begin{equation}\label{poznachenya2}
\begin{split}
& {{W}_{1}}\left( x,\alpha  \right)=\frac{2}{{{w}_{1}}\left( x,\alpha  \right)}+\frac{1}{{{w}_{2}}\left( x,\alpha  \right)}+\frac{1}{{{w}_{3}}\left( x,\alpha  \right)}, \\ 
& {{W}_{2}}\left( x,\alpha  \right)=\frac{{{x}^{2}}s}{{{\left( {{w}_{1}}\left( x,\alpha  \right) \right)}^{2}}}.
\end{split}
\end{equation}

The further calculation of \eqref{tensorposlevika6} depends on the values of indices $a$ and $b$. We demonstrate the calculation workflow for the component $t_{00}^{>}$ only, while the calculation of other components is completely similar. In this case ($a = 0, \; b = 0$) the function $f_{ab}$ in the integrand of \eqref{tensorposlevika6} is equal to ${{f}_{00}} = \left( {{E}^{2}}+{{x}^{2}}s \right)$, so the expression for $t_{00}^>$ has the form
\begin{widetext}
\begin{equation}\label{tensorposlevika7}
\begin{split}
& {{t}_{00}^{>}}=
i
\int\limits_{0}^{\frac{1}{cP}}{xdx}
\int\limits_{0}^{2\pi }{d\alpha }
\left[\vphantom{\int\limits_a^b}
\frac{1}{{{\left( {{w}_{1}}\left( x,\alpha  \right) \right)}^{2}}{{w}_{2}}\left( x,\alpha  \right){{w}_{3}}\left( x,\alpha  \right)}
\right.
 \\ & \times 
 \left.
 \int\limits_{-\infty }^{\infty }{dE}\int\limits_{-\infty }^{\infty }{dy}\left( {{E}^{2}}+{{x}^{2}}s \right)\frac{1}{1+{{W}_{1}}\left( x,\alpha  \right)\left( {{E}^{2}}+{{y}^{2}} \right)+{{W}_{2}}\left( x,\alpha  \right){{E}^{2}}}
\right] 
 .
\end{split}
\end{equation}
\noindent
The integration with respect to  $E$ and $y$ can be reduced to the calculation of the Poisson integrals
\begin{equation}\label{puason}
\begin{split}
& \int\limits_{-\infty }^{\infty }{dE}\int\limits_{-\infty }^{\infty }{dy}\left( {{E}^{2}}+{{x}^{2}}s \right)\frac{1}{1+{{W}_{1}}\left( x,\alpha  \right)\left( {{E}^{2}}+{{y}^{2}} \right)+{{W}_{2}}\left( x,\alpha  \right){{E}^{2}}}= \\ 
& =\frac{\pi }{\sqrt{W\left( x,\alpha  \right)}\sqrt{{{W}_{1}}\left( x,\alpha  \right)}}\left( \frac{1}{2}\frac{1}{\left| W\left( x,\alpha  \right) \right|}+{{x}^{2}}s \right),
\end{split}
\end{equation}
\noindent
where $W\left( x,\alpha  \right) = {{W}_{1}}\left( x,\alpha  \right)+{{W}_{2}}\left( x,\alpha  \right)$. Substituting  \eqref{puason} into \eqref{tensorposlevika7} yields
\begin{equation}\label{tensorposlevika8}
\begin{split}
& {{t}_{00}}^{>}=\pi i\int\limits_{0}^{\frac{1}{cP}}{xdx}\int\limits_{0}^{2\pi }{d\alpha }\frac{1}{{{\left( {{w}_{1}}\left( x,\alpha  \right) \right)}^{2}}{{w}_{2}}\left( x,\alpha  \right){{w}_{3}}\left( x,\alpha  \right)}\times  \\ 
& \times \frac{1}{\sqrt{{{W}_{3}}\left( x,\alpha  \right)}\sqrt{{{W}_{1}}\left( x,\alpha  \right)}}\left( \frac{1}{2}\frac{1}{\left| {{W}_{3}}\left( x,\alpha  \right) \right|}+{{x}^{2}}s \right).  
\end{split}
\end{equation}
\end{widetext}

Finally, we obtain the two-dimensional integral with finite limits which can be calculated numerically. The other non zero components of $t_{ab}^{>}$ have to be calculated in the same way. 

After the components $t^<_{ab}$ and $t^>_{ab}$ get calculated, we can finally calculate their sum $t_{ab} = t^<_{ab} + t^>_{ab}$ and return to the analytical expression \eqref{vyraz_diagramma33} for the diagram \fref{fig:simplest_loops}a. 

So far, we have calculated the analytical expression for the simplest one-loop diagram shown in \fref{fig:simplest_loops}a. However, we should also consider all possible contractions of bispinor and glueball field operators. In the simplest case of the elastic scattering of two protons there is another diagram \fref{fig:simplest_loops}b added to the one-loop diagram \fref{fig:simplest_loops}a. The analytical expression for this diagram is pretty similar to \eqref{vyraz_diagramma33} and the integral in this expression can be calculated in the same way as for $t_{ab}$.

\section{Results and discussion}

\begin{figure*}
	\centering
	\includegraphics[width=0.74\linewidth]{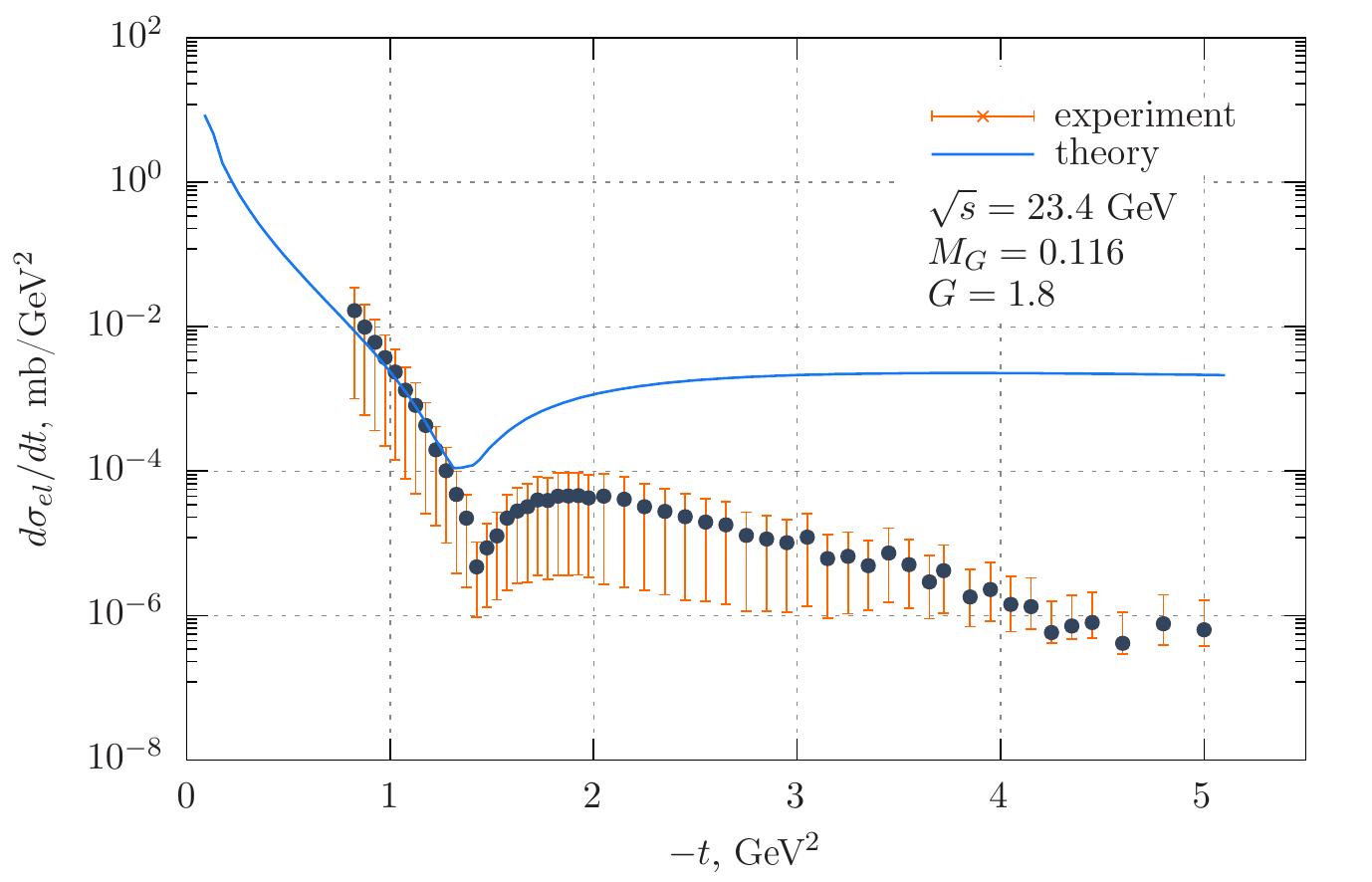}
	\caption[Рис.\ref{fig:dif_sech23}]{
The differential cross-section $d\sigma_{el}/dt\left( t \right)$ of elastic $pp$ scattering at $\sqrt{s}=23.4$ GeV. The solid curve is the calculated dependency ($M_G=0.116$ and $G = 1.8$), the dots with error bars are experimental data \cite{NAGY1979221}.}
	\label{fig:dif_sech23}
\end{figure*}

\begin{figure*}
	\centering
	\includegraphics[width=0.74\linewidth]{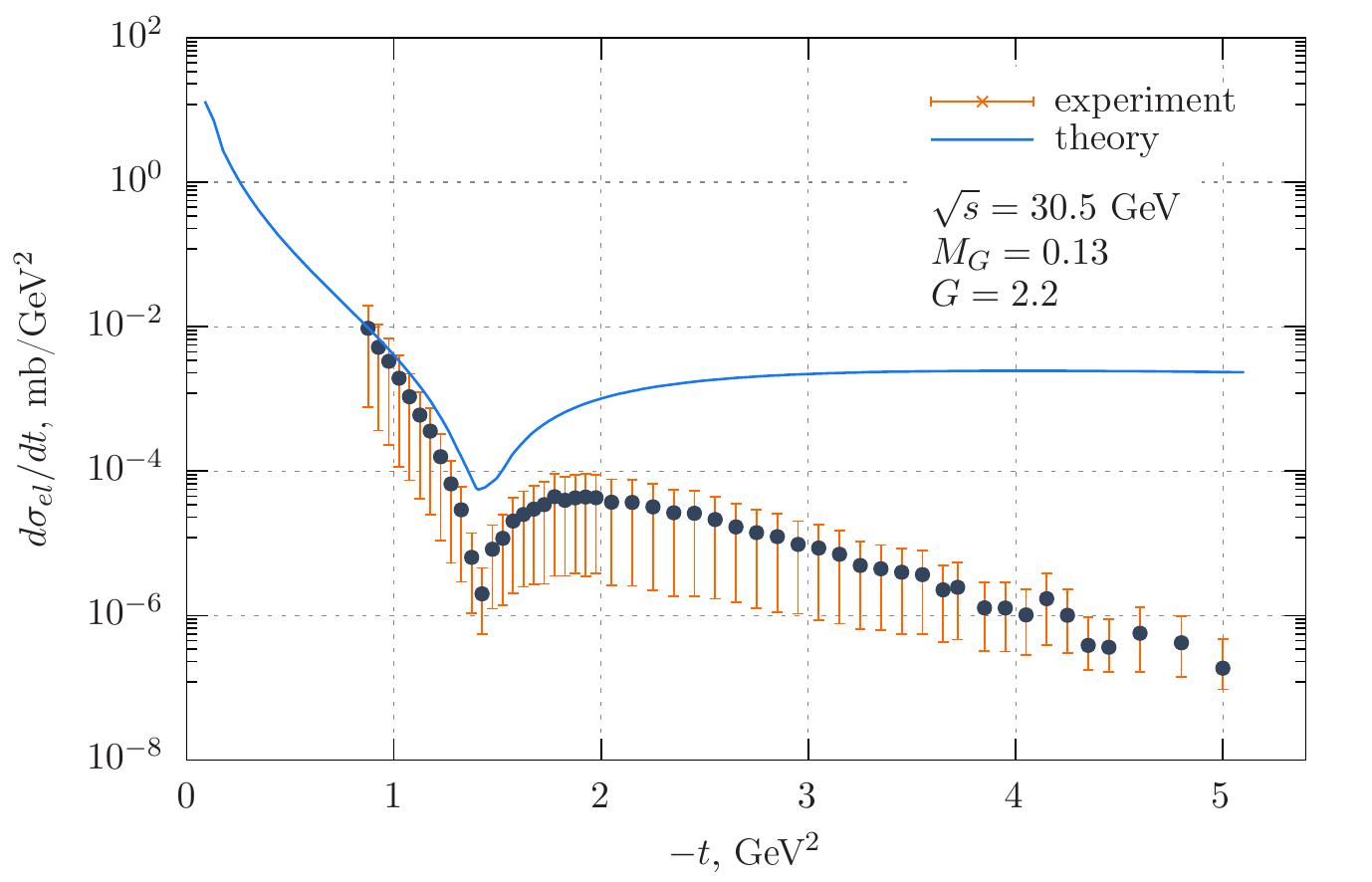}
	\caption[Рис.\ref{fig:dif_sech30}]{
	The differential cross-section $d\sigma_{el}/dt\left( t \right)$  of elastic $pp$ scattering at $\sqrt{s}=30.5$ GeV. The solid  curve is the calculated dependency ($M_G=0.13$ and $G = 2.2$), the dots with error bars are experimental data  \cite{NAGY1979221}.}
	\label{fig:dif_sech30}
\end{figure*}

\begin{figure*}
	\centering
	\includegraphics[width=0.74\linewidth]{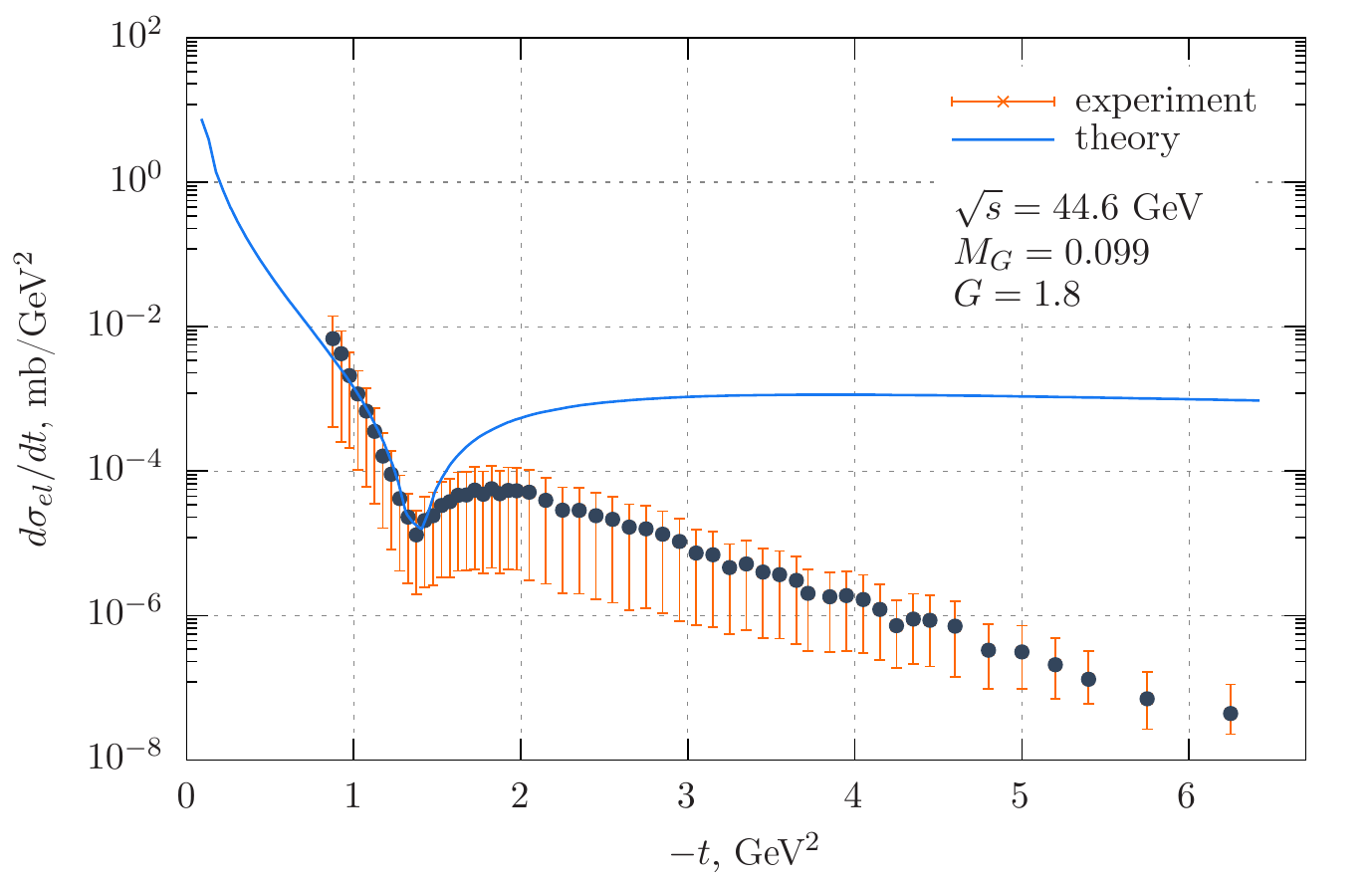}
	\caption[Рис.\ref{fig:dif_sech44}]{The differential cross-section $d\sigma_{el}/dt\left( t \right)$  of elastic $pp$ scattering at $\sqrt{s}=44.6$ GeV. The solid curve is the calculated dependency ($M_G=0.099$ and $G = 1.8$), the dots with error bars are experimental data  \cite{NAGY1979221}.}
	\label{fig:dif_sech44}
\end{figure*}

\begin{figure*}
	\centering
	\includegraphics[width=0.74\linewidth]{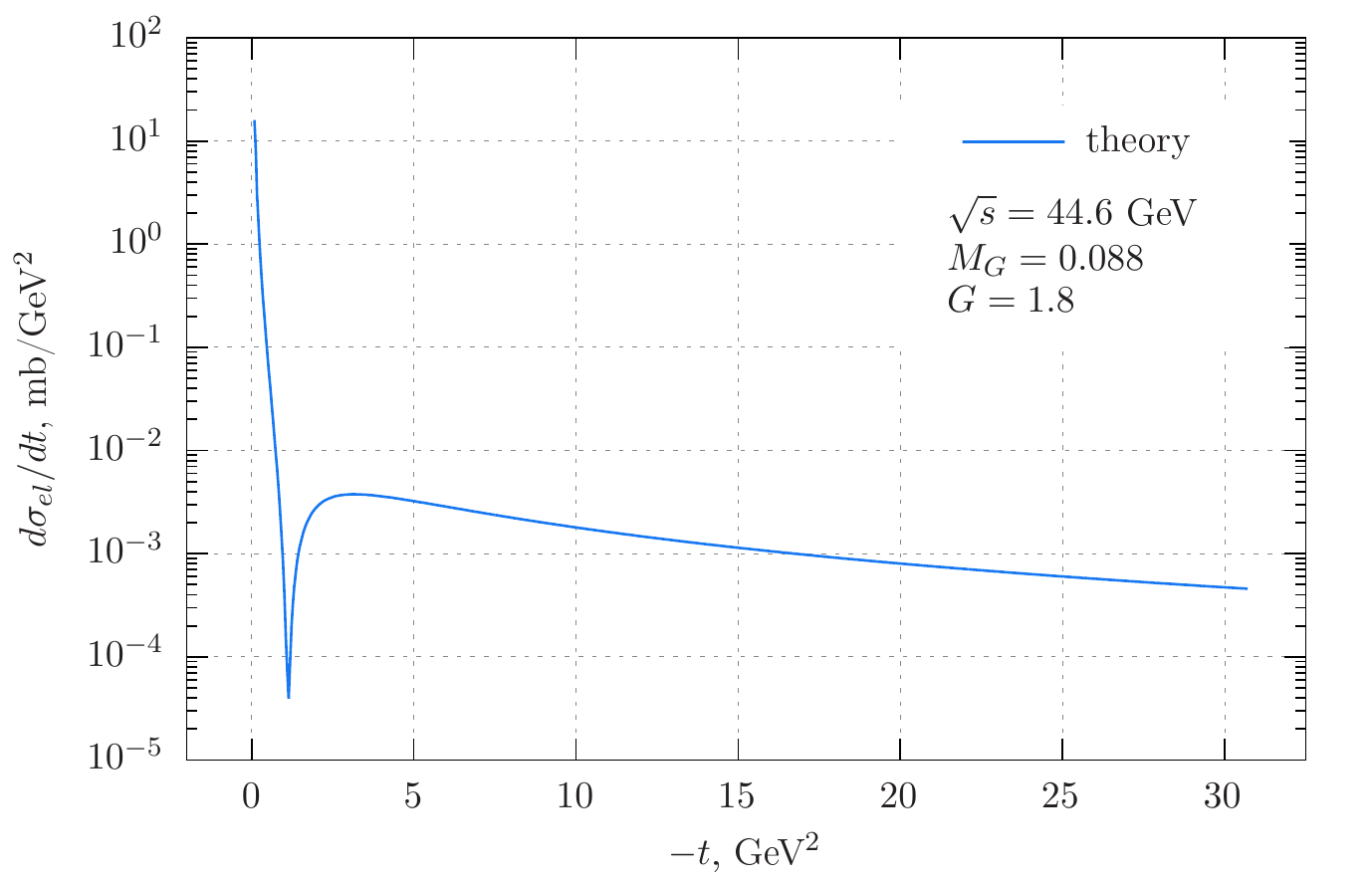}
	\caption{
	The calculated dependency of differential cross-section $d\sigma_{el}/dt\left( t \right)$ of elastic $pp$ scattering at $\sqrt{s}=44.6$ GeV ($M_G=0.088$ and $G = 1.8$). }
	\label{fig:dif_sech_t}
\end{figure*}

The analytic calculations presented in the previous sections allowed us to calculate the differential cross section of elastic proton-proton scattering $d{{\sigma }_{el}}/dt\left( t \right)$. This calculation includes the contributions from the tree-level (pole) (\fref{fig:polusnaya}) and one-loop (\fref{fig:simplest_loops}) diagrams, as well as contributions from the diagrams with the ${{P}_{3}}$ and ${{P}_{4}}$ interchanged. The model used for calculation contains two parameters -- ${{M}_{G}}$ -- the mass of glueball, and $G$ -- the effective coupling of proton-glueball interaction. All the quantities were expressed in the units of the proton mass ${{M}_{P}} = 0.938$ GeV. The obtained dependency $d\sigma_{el}/dt\left( t \right)$ is presented in Fig.\ref{fig:dif_sech23}-\ref{fig:dif_sech_t} at different energies $\sqrt{s}$ and different values of $M_G$ and $G$. 

As can be seen in Fig.\ref{fig:dif_sech23}, the obtained dependency is non-monotonic and qualitatively describes the first minimum of the experimental curve. The values of $M$ and $G$ were chosen in each case (energy) differently to better reproduce the experiment. However, the observed non-monotonic behavior of the theoretical dependency is preserved in all three cases even for the same values of $M$ and $G$.

Although the results obtained are only in qualitative agreement with the experimental data, we expect that inclusion of the diagrams with the higher number of loops may help to achieve the quantitative description of the experiment.

In order to demonstrate the qualitative behavior of the dependency $d\sigma_{el}/dt\left( t \right)$ (including the second fall), we plotted it separately in a wider range of $t$ up to $30$ GeV$^2$ (\fref{fig:dif_sech_t}).

We also performed the analogous calculations within another model $\phi^3$ (phi-cubed) with the scalar field. However, within that model we have not obtained the effects of non-monotonic $d\sigma_{el}/dt\left( t \right)$ dependencies similar to the presented here. It suggests that the physical mechanisms responsible for this non-monotonicity are associated with the spin effects. The spin flip effects observed in the protons scattering have been calculated within the Regge theory \cite{Bence:2020usl}. However, our results are different in that within the Regge approach this non-monotonicity arises due to an exchange of reggeons having different signatures, i.e. the reggeon contributions to the scattering amplitude with different signs. In \cite{Bence:2020usl} the spin effects provides only quantitative changes in the behavior of differential cross-section dependency $d\sigma_{el}/dt\left( t \right)$.  In our approach it is the spin effects that are responsible for the appearance of  non-monotonicity in the differential cross-section dependency $d\sigma_{el}/dt\left( t \right)$ .

\section{Summary}

We demonstrated that the multi-particle fields approach can be used for a development of dynamic models which describe the scattering of multi-quark systems. Using this approach, we considered the elastic proton-proton scattering and calculated the differential cross-section. As a result, we succeeded to obtain the qualitative agreement with the experiment.

Laplace's method allowed us to approximately calculate the loop diagrams. Unfortunately, the calculation is lengthy enough and requires a further improvement.

We found that the experimentally observed non-monotonicity of the differential cross-section dependency $d\sigma_{el}/dt\left( t \right)$ is caused in our model by the spin effects. Taking these effects into account in the tree-level and simplest loop diagrams  allowed us to obtain the qualitative description of experiment. The obtained results suggest that inclusion of the more complex loop diagrams to the calculation will help achieve the quantitative description of experimental data.

\section*{Acknowledgments}
This research is supported by the Ministry of Education and Science of Ukraine (Project No. 0119U103982).

\bibliography{references-utf8}

\begin{thebibliography}{51}
\expandafter\ifx\csname natexlab\endcsname\relax\def\natexlab#1{#1}\fi
\expandafter\ifx\csname bibnamefont\endcsname\relax
  \def\bibnamefont#1{#1}\fi
\expandafter\ifx\csname bibfnamefont\endcsname\relax
  \def\bibfnamefont#1{#1}\fi
\expandafter\ifx\csname citenamefont\endcsname\relax
  \def\citenamefont#1{#1}\fi
\expandafter\ifx\csname url\endcsname\relax
  \def\url#1{\texttt{#1}}\fi
\expandafter\ifx\csname urlprefix\endcsname\relax\def\urlprefix{URL }\fi
\providecommand{\bibinfo}[2]{#2}
\providecommand{\eprint}[2][]{\url{#2}}

\bibitem[{\citenamefont{Cafagna}(2020)}]{Cafagna:2020izf}
\bibinfo{author}{\bibfnamefont{F.}~\bibnamefont{Cafagna}}
  (\bibinfo{collaboration}{TOTEM}), \bibinfo{journal}{PoS}
  \textbf{\bibinfo{volume}{ICRC2019}}, \bibinfo{pages}{207}
  (\bibinfo{year}{2020}).

\bibitem[{\citenamefont{Dremin}(2017)}]{Dremin:2017eng}
\bibinfo{author}{\bibfnamefont{I.~M.} \bibnamefont{Dremin}},
  \bibinfo{journal}{Usp. Fiz. Nauk} \textbf{\bibinfo{volume}{187}},
  \bibinfo{pages}{353} (\bibinfo{year}{2017}),
  \urlprefix\url{https://ufn.ru/ru/articles/2017/4/a/}.

\bibitem[{\citenamefont{Cs\"org\"o et~al.}(2020)\citenamefont{Cs\"org\"o,
  Pasechnik, and Ster}}]{Csorgo:2018ruk}
\bibinfo{author}{\bibfnamefont{T.}~\bibnamefont{Cs\"org\"o}},
  \bibinfo{author}{\bibfnamefont{R.}~\bibnamefont{Pasechnik}},
  \bibnamefont{and} \bibinfo{author}{\bibfnamefont{A.}~\bibnamefont{Ster}},
  \bibinfo{journal}{Phys. Part. Nucl.} \textbf{\bibinfo{volume}{51}},
  \bibinfo{pages}{227} (\bibinfo{year}{2020}), \eprint{1811.08913}.

\bibitem[{\citenamefont{Collins}(1977)}]{Collins1977}
\bibinfo{author}{\bibfnamefont{P.~D.~B.} \bibnamefont{Collins}},
  \emph{\bibinfo{title}{An Introduction to Regge Theory and High Energy
  Physics}} (\bibinfo{publisher}{Cambridge University Press},
  \bibinfo{year}{1977}).

\bibitem[{\citenamefont{Ryskin et~al.}(2009)\citenamefont{Ryskin, Martin,
  Khoze, and Shuvaev}}]{Ryskin:2009qf}
\bibinfo{author}{\bibfnamefont{M.}~\bibnamefont{Ryskin}},
  \bibinfo{author}{\bibfnamefont{A.}~\bibnamefont{Martin}},
  \bibinfo{author}{\bibfnamefont{V.}~\bibnamefont{Khoze}}, \bibnamefont{and}
  \bibinfo{author}{\bibfnamefont{A.}~\bibnamefont{Shuvaev}},
  \bibinfo{journal}{J. Phys. G} \textbf{\bibinfo{volume}{36}},
  \bibinfo{pages}{093001} (\bibinfo{year}{2009}), \eprint{0907.1374}.

\bibitem[{\citenamefont{Dremin}(2019)}]{Dremin:2019pza}
\bibinfo{author}{\bibfnamefont{I.}~\bibnamefont{Dremin}},
  \bibinfo{journal}{Particles} \textbf{\bibinfo{volume}{2}},
  \bibinfo{pages}{57} (\bibinfo{year}{2019}).

\bibitem[{\citenamefont{Islam and Luddy}(2017)}]{Islam:2017muy}
\bibinfo{author}{\bibfnamefont{M.}~\bibnamefont{Islam}} \bibnamefont{and}
  \bibinfo{author}{\bibfnamefont{R.}~\bibnamefont{Luddy}}, in
  \emph{\bibinfo{booktitle}{{17th conference on Elastic and Diffractive
  Scattering}}} (\bibinfo{year}{2017}), \eprint{1709.03172}.

\bibitem[{\citenamefont{Amaldi}(2015)}]{UgoAmaldi}
\bibinfo{author}{\bibfnamefont{U.}~\bibnamefont{Amaldi}},
  \emph{\bibinfo{title}{An ISR Discovery: The Rise of the Proton–Proton
  Cross-Section}} (\bibinfo{year}{2015}), pp. \bibinfo{pages}{257--286}.

\bibitem[{\citenamefont{Fadin et~al.}(1975)\citenamefont{Fadin, Kuraev, and
  Lipatov}}]{FADIN197550}
\bibinfo{author}{\bibfnamefont{V.}~\bibnamefont{Fadin}},
  \bibinfo{author}{\bibfnamefont{E.}~\bibnamefont{Kuraev}}, \bibnamefont{and}
  \bibinfo{author}{\bibfnamefont{L.}~\bibnamefont{Lipatov}},
  \bibinfo{journal}{Physics Letters B} \textbf{\bibinfo{volume}{60}},
  \bibinfo{pages}{50 } (\bibinfo{year}{1975}), ISSN \bibinfo{issn}{0370-2693},
  \urlprefix\url{http://www.sciencedirect.com/science/article/pii/0370269375905249}.

\bibitem[{\citenamefont{Kuraev et~al.}(1977)\citenamefont{Kuraev, Lipatov, and
  Fadin}}]{Kuraev:1977eng}
\bibinfo{author}{\bibfnamefont{E.}~\bibnamefont{Kuraev}},
  \bibinfo{author}{\bibfnamefont{L.}~\bibnamefont{Lipatov}}, \bibnamefont{and}
  \bibinfo{author}{\bibfnamefont{V.}~\bibnamefont{Fadin}},
  \bibinfo{journal}{Sov. Phys. - JETP.} \textbf{\bibinfo{volume}{72}},
  \bibinfo{pages}{377} (\bibinfo{year}{1977}).

\bibitem[{\citenamefont{Lipatov}(2008)}]{Lipatov:2008eng}
\bibinfo{author}{\bibfnamefont{L.~N.} \bibnamefont{Lipatov}},
  \bibinfo{journal}{Usp. Fiz. Nauk} \textbf{\bibinfo{volume}{178}},
  \bibinfo{pages}{663} (\bibinfo{year}{2008}),
  \urlprefix\url{https://ufn.ru/ru/articles/2008/6/m/}.

\bibitem[{\citenamefont{Kozlov et~al.}(2007)\citenamefont{Kozlov, Reznichenko,
  and Fadin}}]{KRFeng}
\bibinfo{author}{\bibfnamefont{M.~G.} \bibnamefont{Kozlov}},
  \bibinfo{author}{\bibfnamefont{A.~V.} \bibnamefont{Reznichenko}},
  \bibnamefont{and} \bibinfo{author}{\bibfnamefont{V.~S.} \bibnamefont{Fadin}},
  \textbf{\bibinfo{volume}{2}}, \bibinfo{pages}{3} (\bibinfo{year}{2007}), ISSN
  \bibinfo{issn}{1818-7994}.

\bibitem[{\citenamefont{Kuraev et~al.}(1976)\citenamefont{Kuraev, Lipatov, and
  Fadin}}]{MultiRggeBFKLeng}
\bibinfo{author}{\bibfnamefont{E.}~\bibnamefont{Kuraev}},
  \bibinfo{author}{\bibfnamefont{L.}~\bibnamefont{Lipatov}}, \bibnamefont{and}
  \bibinfo{author}{\bibfnamefont{V.}~\bibnamefont{Fadin}},
  \bibinfo{journal}{Sov. Phys. JETP} \textbf{\bibinfo{volume}{44}},
  \bibinfo{pages}{443} (\bibinfo{year}{1976}).

\bibitem[{\citenamefont{Duhr and Liu}(2019)}]{Duhr2019}
\bibinfo{author}{\bibfnamefont{C.}~\bibnamefont{Duhr}} \bibnamefont{and}
  \bibinfo{author}{\bibfnamefont{Z.}~\bibnamefont{Liu}},
  \bibinfo{journal}{Journal of High Energy Physics}
  \textbf{\bibinfo{volume}{2019}}, \bibinfo{pages}{146} (\bibinfo{year}{2019}),
  ISSN \bibinfo{issn}{1029-8479},
  \urlprefix\url{https://doi.org/10.1007/JHEP01(2019)146}.

\bibitem[{\citenamefont{Hentschinski}(2020)}]{Hentschinski:2020rfx}
\bibinfo{author}{\bibfnamefont{M.}~\bibnamefont{Hentschinski}}
  (\bibinfo{year}{2020}), \eprint{2010.14748}.

\bibitem[{\citenamefont{Sharph et~al.}(2012)\citenamefont{Sharph, Tykhonov,
  Sokhrannyi, Deliyergiyev, Podolyan et~al.}}]{Sharph:2011wm}
\bibinfo{author}{\bibfnamefont{I.}~\bibnamefont{Sharph}},
  \bibinfo{author}{\bibfnamefont{A.}~\bibnamefont{Tykhonov}},
  \bibinfo{author}{\bibfnamefont{G.}~\bibnamefont{Sokhrannyi}},
  \bibinfo{author}{\bibfnamefont{M.}~\bibnamefont{Deliyergiyev}},
  \bibinfo{author}{\bibfnamefont{N.}~\bibnamefont{Podolyan}},
  \bibnamefont{et~al.}, \bibinfo{journal}{Central Eur.J.Phys.}
  \textbf{\bibinfo{volume}{10}}, \bibinfo{pages}{858} (\bibinfo{year}{2012}),
  \eprint{1110.4945}.

\bibitem[{\citenamefont{Dremin}(2013)}]{Dremin:2013eng}
\bibinfo{author}{\bibfnamefont{I.~M.} \bibnamefont{Dremin}},
  \bibinfo{journal}{Phys. Usp.} \textbf{\bibinfo{volume}{56}},
  \bibinfo{pages}{3} (\bibinfo{year}{2013}),
  \urlprefix\url{https://ufn.ru/en/articles/2013/1/b/}.

\bibitem[{\citenamefont{Abramovsky et~al.}(1973)\citenamefont{Abramovsky,
  Gribov, and Kancheli}}]{AGKeng}
\bibinfo{author}{\bibfnamefont{V.~A.} \bibnamefont{Abramovsky}},
  \bibinfo{author}{\bibfnamefont{V.~N.} \bibnamefont{Gribov}},
  \bibnamefont{and} \bibinfo{author}{\bibfnamefont{O.~V.}
  \bibnamefont{Kancheli}}, \bibinfo{journal}{Sov. J. Nucl. Phys.}
  \textbf{\bibinfo{volume}{18}}, \bibinfo{pages}{308} (\bibinfo{year}{1973}).

\bibitem[{\citenamefont{Levin and Prygarin}(2008)}]{Levin:1900tt}
\bibinfo{author}{\bibfnamefont{E.}~\bibnamefont{Levin}} \bibnamefont{and}
  \bibinfo{author}{\bibfnamefont{A.}~\bibnamefont{Prygarin}},
  \bibinfo{journal}{Phys. Rev. C} \textbf{\bibinfo{volume}{78}},
  \bibinfo{pages}{065202} (\bibinfo{year}{2008}), \eprint{0804.4747}.

\bibitem[{\citenamefont{Bartels et~al.}(2008)\citenamefont{Bartels, Salvadore,
  and Vacca}}]{Bartels:2008ru}
\bibinfo{author}{\bibfnamefont{J.}~\bibnamefont{Bartels}},
  \bibinfo{author}{\bibfnamefont{M.}~\bibnamefont{Salvadore}},
  \bibnamefont{and} \bibinfo{author}{\bibfnamefont{G.}~\bibnamefont{Vacca}},
  \bibinfo{journal}{JHEP} \textbf{\bibinfo{volume}{06}}, \bibinfo{pages}{032}
  (\bibinfo{year}{2008}), \eprint{0802.2702}.

\bibitem[{\citenamefont{Pelicer et~al.}(2019)\citenamefont{Pelicer,
  de~Oliveira, Martin, and Ryskin}}]{Pelicer:2018xzu}
\bibinfo{author}{\bibfnamefont{M.}~\bibnamefont{Pelicer}},
  \bibinfo{author}{\bibfnamefont{E.}~\bibnamefont{de~Oliveira}},
  \bibinfo{author}{\bibfnamefont{A.}~\bibnamefont{Martin}}, \bibnamefont{and}
  \bibinfo{author}{\bibfnamefont{M.}~\bibnamefont{Ryskin}},
  \bibinfo{journal}{Eur. Phys. J. C} \textbf{\bibinfo{volume}{79}},
  \bibinfo{pages}{9} (\bibinfo{year}{2019}), \eprint{1810.05573}.

\bibitem[{\citenamefont{Gribov}()}]{Gribov_Reggeonnaja_technica_eng}
\bibinfo{author}{\bibfnamefont{V.~N.} \bibnamefont{Gribov}},
  \bibinfo{journal}{Sov. Phys. JETP} \textbf{\bibinfo{volume}{26}},
  \bibinfo{pages}{414} (????).

\bibitem[{\citenamefont{Baker and Ter-Martirosyan}(1976)}]{BAKER19761}
\bibinfo{author}{\bibfnamefont{M.}~\bibnamefont{Baker}} \bibnamefont{and}
  \bibinfo{author}{\bibfnamefont{K.}~\bibnamefont{Ter-Martirosyan}},
  \bibinfo{journal}{Physics Reports} \textbf{\bibinfo{volume}{28}},
  \bibinfo{pages}{1 } (\bibinfo{year}{1976}), ISSN \bibinfo{issn}{0370-1573},
  \urlprefix\url{http://www.sciencedirect.com/science/article/pii/0370157376900028}.

\bibitem[{\citenamefont{Yu. P.~Nikitin}(1980)}]{Nikitin1980}
\bibinfo{author}{\bibfnamefont{I.~L.~R.} \bibnamefont{Yu. P.~Nikitin}},
  \emph{\bibinfo{title}{High energy nuclear physics}}
  (\bibinfo{publisher}{Atomizdat}, \bibinfo{year}{1980}), \bibinfo{note}{in
  Russian}.

\bibitem[{\citenamefont{Shabelski and Shuvaev}(2014)}]{Shabelski:2014yba}
\bibinfo{author}{\bibfnamefont{Y.~M.} \bibnamefont{Shabelski}}
  \bibnamefont{and} \bibinfo{author}{\bibfnamefont{A.}~\bibnamefont{Shuvaev}},
  \bibinfo{journal}{JHEP} \textbf{\bibinfo{volume}{11}}, \bibinfo{pages}{023}
  (\bibinfo{year}{2014}), \eprint{1406.1421}.

\bibitem[{\citenamefont{Kaidalov}(2003)}]{Kaidalov:2003eng}
\bibinfo{author}{\bibfnamefont{A.~B.} \bibnamefont{Kaidalov}},
  \bibinfo{journal}{Usp. Fiz. Nauk} \textbf{\bibinfo{volume}{173}},
  \bibinfo{pages}{1153} (\bibinfo{year}{2003}),
  \urlprefix\url{https://ufn.ru/ru/articles/2003/11/a/}.

\bibitem[{\citenamefont{Ter-Martirosyan}(1975)}]{Ter-Martirosyan}
\bibinfo{author}{\bibfnamefont{K.}~\bibnamefont{Ter-Martirosyan}},
  \emph{\bibinfo{title}{Results of Regge scheme development and experiment}}
  (\bibinfo{publisher}{MIPHI}, \bibinfo{address}{Moscow},
  \bibinfo{year}{1975}).

\bibitem[{\citenamefont{Nagy et~al.}(1979)}]{NAGY1979221}
\bibinfo{author}{\bibfnamefont{E.}~\bibnamefont{Nagy}} \bibnamefont{et~al.},
  \bibinfo{journal}{Nuclear Physics B} \textbf{\bibinfo{volume}{150}},
  \bibinfo{pages}{221} (\bibinfo{year}{1979}).

\bibitem[{\citenamefont{Breakstone et~al.}(1985)}]{PhysRevLett.54.2180}
\bibinfo{author}{\bibfnamefont{A.}~\bibnamefont{Breakstone}}
  \bibnamefont{et~al.}, \bibinfo{journal}{Phys. Rev. Lett.}
  \textbf{\bibinfo{volume}{54}}, \bibinfo{pages}{2180} (\bibinfo{year}{1985}),
  \urlprefix\url{https://link.aps.org/doi/10.1103/PhysRevLett.54.2180}.

\bibitem[{\citenamefont{Antchev et~al.}(2019)}]{Antchev:2018edk}
\bibinfo{author}{\bibfnamefont{G.}~\bibnamefont{Antchev}} \bibnamefont{et~al.}
  (\bibinfo{collaboration}{TOTEM}), \bibinfo{journal}{Eur. Phys. J. C}
  \textbf{\bibinfo{volume}{79}}, \bibinfo{pages}{861} (\bibinfo{year}{2019}),
  \eprint{1812.08283}.

\bibitem[{\citenamefont{Abazov et~al.}(2020)}]{Abazov:2020rus}
\bibinfo{author}{\bibfnamefont{V.}~\bibnamefont{Abazov}} \bibnamefont{et~al.}
  (\bibinfo{year}{2020}), \eprint{2012.03981}.

\bibitem[{\citenamefont{Kaidalov and
  Ter-Martirosian}(1984)}]{Kaidalov:1983vn_eng}
\bibinfo{author}{\bibfnamefont{A.~B.} \bibnamefont{Kaidalov}} \bibnamefont{and}
  \bibinfo{author}{\bibfnamefont{K.~A.} \bibnamefont{Ter-Martirosian}},
  \bibinfo{journal}{Sov. J. Nucl. Phys.} \textbf{\bibinfo{volume}{39}},
  \bibinfo{pages}{979} (\bibinfo{year}{1984}).

\bibitem[{\citenamefont{Kaidalov and Ter-Martirosyan}(1984)}]{Kaidalov:1983ew}
\bibinfo{author}{\bibfnamefont{A.}~\bibnamefont{Kaidalov}} \bibnamefont{and}
  \bibinfo{author}{\bibfnamefont{K.}~\bibnamefont{Ter-Martirosyan}},
  \bibinfo{journal}{Sov. J. Nucl. Phys.} \textbf{\bibinfo{volume}{40}},
  \bibinfo{pages}{135} (\bibinfo{year}{1984}).

\bibitem[{\citenamefont{Vall et~al.}(1988)\citenamefont{Vall, Jenkovszky, and
  Struminsky}}]{Dipol_eng}
\bibinfo{author}{\bibfnamefont{A.}~\bibnamefont{Vall}},
  \bibinfo{author}{\bibfnamefont{L.}~\bibnamefont{Jenkovszky}},
  \bibnamefont{and}
  \bibinfo{author}{\bibfnamefont{B.}~\bibnamefont{Struminsky}},
  \bibinfo{journal}{Sov. J. Part. Nucl.} \textbf{\bibinfo{volume}{19}},
  \bibinfo{pages}{181} (\bibinfo{year}{1988}).

\bibitem[{\citenamefont{Martynov}(2013)}]{PhysRevD.87.114018}
\bibinfo{author}{\bibfnamefont{E.}~\bibnamefont{Martynov}},
  \bibinfo{journal}{Phys. Rev. D} \textbf{\bibinfo{volume}{87}},
  \bibinfo{pages}{114018} (\bibinfo{year}{2013}),
  \urlprefix\url{https://link.aps.org/doi/10.1103/PhysRevD.87.114018}.

\bibitem[{\citenamefont{Martynov}(2007)}]{PhysRevD.76.074030}
\bibinfo{author}{\bibfnamefont{E.}~\bibnamefont{Martynov}},
  \bibinfo{journal}{Phys. Rev. D} \textbf{\bibinfo{volume}{76}},
  \bibinfo{pages}{074030} (\bibinfo{year}{2007}),
  \urlprefix\url{https://link.aps.org/doi/10.1103/PhysRevD.76.074030}.

\bibitem[{\citenamefont{{Martynov, Evgenij} and {Nicolescu,
  Basarab}}(2019)}]{refId0}
\bibinfo{author}{\bibnamefont{{Martynov, Evgenij}}} \bibnamefont{and}
  \bibinfo{author}{\bibnamefont{{Nicolescu, Basarab}}}, \bibinfo{journal}{Eur.
  Phys. J. C} \textbf{\bibinfo{volume}{79}}, \bibinfo{pages}{461}
  (\bibinfo{year}{2019}),
  \urlprefix\url{https://doi.org/10.1140/epjc/s10052-019-6954-6}.

\bibitem[{\citenamefont{Ptashynskiy
  et~al.}(2019{\natexlab{a}})\citenamefont{Ptashynskiy, Zelentsova, Chudak,
  Merkotan, Potiienko, Voitenko, Berezovskiy, Opyatyuk, Zharova, Yushkevich
  et~al.}}]{Korotca_statta_v_UJP}
\bibinfo{author}{\bibfnamefont{D.}~\bibnamefont{Ptashynskiy}},
  \bibinfo{author}{\bibfnamefont{T.}~\bibnamefont{Zelentsova}},
  \bibinfo{author}{\bibfnamefont{N.}~\bibnamefont{Chudak}},
  \bibinfo{author}{\bibfnamefont{K.}~\bibnamefont{Merkotan}},
  \bibinfo{author}{\bibfnamefont{O.}~\bibnamefont{Potiienko}},
  \bibinfo{author}{\bibfnamefont{V.}~\bibnamefont{Voitenko}},
  \bibinfo{author}{\bibfnamefont{O.}~\bibnamefont{Berezovskiy}},
  \bibinfo{author}{\bibfnamefont{V.}~\bibnamefont{Opyatyuk}},
  \bibinfo{author}{\bibfnamefont{O.}~\bibnamefont{Zharova}},
  \bibinfo{author}{\bibfnamefont{T.}~\bibnamefont{Yushkevich}},
  \bibnamefont{et~al.}, \bibinfo{journal}{Ukrainian Journal of Physics}
  \textbf{\bibinfo{volume}{64}}, \bibinfo{pages}{732}
  (\bibinfo{year}{2019}{\natexlab{a}}),
  \urlprefix\url{https://ujp.bitp.kiev.ua/index.php/ujp/article/view/2019394}.

\bibitem[{\citenamefont{Ptashynskiy
  et~al.}(2019{\natexlab{b}})\citenamefont{Ptashynskiy, Zelentsova, Chudak,
  Merkotan, Potiienko, Voitenko, Berezovskiy, Opyatyuk, Zharova, Yushkevich
  et~al.}}]{Ptashynskiy2019MultiparticleFO}
\bibinfo{author}{\bibfnamefont{D.~A.} \bibnamefont{Ptashynskiy}},
  \bibinfo{author}{\bibfnamefont{T.~M.} \bibnamefont{Zelentsova}},
  \bibinfo{author}{\bibfnamefont{N.~O.} \bibnamefont{Chudak}},
  \bibinfo{author}{\bibfnamefont{K.~K.} \bibnamefont{Merkotan}},
  \bibinfo{author}{\bibfnamefont{O.~S.} \bibnamefont{Potiienko}},
  \bibinfo{author}{\bibfnamefont{V.~V.} \bibnamefont{Voitenko}},
  \bibinfo{author}{\bibfnamefont{O.~D.} \bibnamefont{Berezovskiy}},
  \bibinfo{author}{\bibfnamefont{V.~V.} \bibnamefont{Opyatyuk}},
  \bibinfo{author}{\bibfnamefont{O.~V.} \bibnamefont{Zharova}},
  \bibinfo{author}{\bibfnamefont{T.~V.} \bibnamefont{Yushkevich}},
  \bibnamefont{et~al.}, \bibinfo{journal}{arXiv: General Physics}
  (\bibinfo{year}{2019}{\natexlab{b}}).

\bibitem[{\citenamefont{Potiienko
  et~al.}(2020{\natexlab{a}})\citenamefont{Potiienko, Sharph, Zelentsova,
  Chudak, Neboga, Merkotan, and Ptashynskiy}}]{Sharph:2020gki}
\bibinfo{author}{\bibfnamefont{O.}~\bibnamefont{Potiienko}},
  \bibinfo{author}{\bibfnamefont{I.}~\bibnamefont{Sharph}},
  \bibinfo{author}{\bibfnamefont{T.}~\bibnamefont{Zelentsova}},
  \bibinfo{author}{\bibfnamefont{N.}~\bibnamefont{Chudak}},
  \bibinfo{author}{\bibfnamefont{G.}~\bibnamefont{Neboga}},
  \bibinfo{author}{\bibfnamefont{K.}~\bibnamefont{Merkotan}}, \bibnamefont{and}
  \bibinfo{author}{\bibfnamefont{D.}~\bibnamefont{Ptashynskiy}}
  (\bibinfo{year}{2020}{\natexlab{a}}), \eprint{2010.06679}.

\bibitem[{\citenamefont{Cartiglia}(2013)}]{Cartiglia:2013jya}
\bibinfo{author}{\bibfnamefont{N.}~\bibnamefont{Cartiglia}}, in
  \emph{\bibinfo{booktitle}{{32nd International Symposium on Physics in
  Collision}}} (\bibinfo{year}{2013}), pp. \bibinfo{pages}{55--64},
  \eprint{1303.2927}.

\bibitem[{\citenamefont{Potiienko
  et~al.}(2020{\natexlab{b}})\citenamefont{Potiienko, Merkotan, Chudak,
  Ptashynskiy, Zelentsova, Yushkevich, Sharph, and
  Rusov}}]{PhysRevD.101.076021}
\bibinfo{author}{\bibfnamefont{O.}~\bibnamefont{Potiienko}},
  \bibinfo{author}{\bibfnamefont{K.}~\bibnamefont{Merkotan}},
  \bibinfo{author}{\bibfnamefont{N.}~\bibnamefont{Chudak}},
  \bibinfo{author}{\bibfnamefont{D.}~\bibnamefont{Ptashynskiy}},
  \bibinfo{author}{\bibfnamefont{T.}~\bibnamefont{Zelentsova}},
  \bibinfo{author}{\bibfnamefont{T.}~\bibnamefont{Yushkevich}},
  \bibinfo{author}{\bibfnamefont{I.}~\bibnamefont{Sharph}}, \bibnamefont{and}
  \bibinfo{author}{\bibfnamefont{V.}~\bibnamefont{Rusov}},
  \bibinfo{journal}{Phys. Rev. D} \textbf{\bibinfo{volume}{101}},
  \bibinfo{pages}{076021} (\bibinfo{year}{2020}{\natexlab{b}}),
  \urlprefix\url{https://link.aps.org/doi/10.1103/PhysRevD.101.076021}.

\bibitem[{\citenamefont{Eden et~al.}(1966)\citenamefont{Eden, Landshoff, Olive,
  and Polkinghorne}}]{Eden:98637}
\bibinfo{author}{\bibfnamefont{R.~J.} \bibnamefont{Eden}},
  \bibinfo{author}{\bibfnamefont{P.~V.} \bibnamefont{Landshoff}},
  \bibinfo{author}{\bibfnamefont{D.~I.} \bibnamefont{Olive}}, \bibnamefont{and}
  \bibinfo{author}{\bibfnamefont{J.~C.} \bibnamefont{Polkinghorne}},
  \emph{\bibinfo{title}{{The analytic S-matrix}}}
  (\bibinfo{publisher}{Cambridge Univ. Press}, \bibinfo{address}{Cambridge},
  \bibinfo{year}{1966}), \urlprefix\url{https://cds.cern.ch/record/98637}.

\bibitem[{\citenamefont{L\'evy and Sucher}(1969)}]{PhysRev.186.1656}
\bibinfo{author}{\bibfnamefont{M.}~\bibnamefont{L\'evy}} \bibnamefont{and}
  \bibinfo{author}{\bibfnamefont{J.}~\bibnamefont{Sucher}},
  \bibinfo{journal}{Phys. Rev.} \textbf{\bibinfo{volume}{186}},
  \bibinfo{pages}{1656} (\bibinfo{year}{1969}),
  \urlprefix\url{https://link.aps.org/doi/10.1103/PhysRev.186.1656}.

\bibitem[{\citenamefont{De~Bruijn}(1958)}]{DeBruijn:225131}
\bibinfo{author}{\bibfnamefont{N.~G.} \bibnamefont{De~Bruijn}},
  \emph{\bibinfo{title}{Asymptotic methods in analysis; 1st ed.}}, Bibl.
  Matematica (\bibinfo{publisher}{North-Holland}, \bibinfo{address}{Amsterdam},
  \bibinfo{year}{1958}).

\bibitem[{\citenamefont{Sharf et~al.}(2011)\citenamefont{Sharf, Tykhonov,
  Sokhrannyi, Yatkin, Deliyergiyev, Podolyan, and Rusov}}]{Sharf:2011ujp}
\bibinfo{author}{\bibfnamefont{I.}~\bibnamefont{Sharf}},
  \bibinfo{author}{\bibfnamefont{A.}~\bibnamefont{Tykhonov}},
  \bibinfo{author}{\bibfnamefont{G.}~\bibnamefont{Sokhrannyi}},
  \bibinfo{author}{\bibfnamefont{K.}~\bibnamefont{Yatkin}},
  \bibinfo{author}{\bibfnamefont{M.}~\bibnamefont{Deliyergiyev}},
  \bibinfo{author}{\bibfnamefont{N.}~\bibnamefont{Podolyan}}, \bibnamefont{and}
  \bibinfo{author}{\bibfnamefont{V.}~\bibnamefont{Rusov}},
  \bibinfo{journal}{Ukr. J. Phys.} \textbf{\bibinfo{volume}{56}},
  \bibinfo{pages}{1151} (\bibinfo{year}{2011}).

\bibitem[{\citenamefont{Sharf and Rusov}(2011)}]{Sharf:2006it}
\bibinfo{author}{\bibfnamefont{I.}~\bibnamefont{Sharf}} \bibnamefont{and}
  \bibinfo{author}{\bibfnamefont{V.}~\bibnamefont{Rusov}},
  \bibinfo{journal}{J.Mod.Phys.} \textbf{\bibinfo{volume}{2}},
  \bibinfo{pages}{1480} (\bibinfo{year}{2011}), \eprint{hep-ph/0605110}.

\bibitem[{\citenamefont{Sharf et~al.}(2012{\natexlab{a}})\citenamefont{Sharf,
  Haj Farajallah~Dabbagh, Tikhonov, and Rusov}}]{Sharf:2007cf}
\bibinfo{author}{\bibfnamefont{I.}~\bibnamefont{Sharf}},
  \bibinfo{author}{\bibfnamefont{A.}~\bibnamefont{Haj Farajallah~Dabbagh}},
  \bibinfo{author}{\bibfnamefont{A.}~\bibnamefont{Tikhonov}}, \bibnamefont{and}
  \bibinfo{author}{\bibfnamefont{V.}~\bibnamefont{Rusov}},
  \bibinfo{journal}{J.Mod.Phys.} \textbf{\bibinfo{volume}{3}},
  \bibinfo{pages}{16} (\bibinfo{year}{2012}{\natexlab{a}}), \eprint{0711.3690}.

\bibitem[{\citenamefont{Sharf et~al.}(2012{\natexlab{b}})\citenamefont{Sharf,
  Tykhonov, Sokhrannyi, Yatkin, and Rusov}}]{Sharf:2009yy}
\bibinfo{author}{\bibfnamefont{I.}~\bibnamefont{Sharf}},
  \bibinfo{author}{\bibfnamefont{A.}~\bibnamefont{Tykhonov}},
  \bibinfo{author}{\bibfnamefont{G.}~\bibnamefont{Sokhrannyi}},
  \bibinfo{author}{\bibfnamefont{K.}~\bibnamefont{Yatkin}}, \bibnamefont{and}
  \bibinfo{author}{\bibfnamefont{V.}~\bibnamefont{Rusov}},
  \bibinfo{journal}{J.Mod.Phys.} \textbf{\bibinfo{volume}{3}},
  \bibinfo{pages}{129} (\bibinfo{year}{2012}{\natexlab{b}}),
  \eprint{0912.2598}.

\bibitem[{\citenamefont{Sharph et~al.}(2020)\citenamefont{Sharph, Zelentsova,
  Chudak, Potiienko, Ptashynskiy, Merkotan, Yushkevich, and
  Milieva}}]{sharph2020laplace}
\bibinfo{author}{\bibfnamefont{I.~V.} \bibnamefont{Sharph}},
  \bibinfo{author}{\bibfnamefont{T.~M.} \bibnamefont{Zelentsova}},
  \bibinfo{author}{\bibfnamefont{N.~O.} \bibnamefont{Chudak}},
  \bibinfo{author}{\bibfnamefont{O.~S.} \bibnamefont{Potiienko}},
  \bibinfo{author}{\bibfnamefont{D.~A.} \bibnamefont{Ptashynskiy}},
  \bibinfo{author}{\bibfnamefont{K.~K.} \bibnamefont{Merkotan}},
  \bibinfo{author}{\bibfnamefont{T.~V.} \bibnamefont{Yushkevich}},
  \bibnamefont{and} \bibinfo{author}{\bibfnamefont{A.~O.}
  \bibnamefont{Milieva}}, \emph{\bibinfo{title}{Laplace method for the simplest
  diagrams of elastic scattering of scalar particles}} (\bibinfo{year}{2020}),
  \eprint{2003.07321}.

\bibitem[{\citenamefont{Bence et~al.}(2020)\citenamefont{Bence, Lengyel,
  Tarics, Martynov, and Tersimonov}}]{Bence:2020usl}
\bibinfo{author}{\bibfnamefont{N.}~\bibnamefont{Bence}},
  \bibinfo{author}{\bibfnamefont{A.}~\bibnamefont{Lengyel}},
  \bibinfo{author}{\bibfnamefont{Z.}~\bibnamefont{Tarics}},
  \bibinfo{author}{\bibfnamefont{E.}~\bibnamefont{Martynov}}, \bibnamefont{and}
  \bibinfo{author}{\bibfnamefont{G.}~\bibnamefont{Tersimonov}}
  (\bibinfo{year}{2020}), \eprint{2010.11987}.

\end{thebibliography}

\end{document}